\documentclass[journal,10pt]{IEEEtran}
\usepackage{amsmath,amssymb,amsfonts}
\usepackage{mathrsfs}
\usepackage{graphicx}
\usepackage{float}
\usepackage{subfig}
\usepackage{color}
\usepackage{cite}
\usepackage{multirow}
\usepackage{algorithm}  
\usepackage{algorithmicx}  
\usepackage{algpseudocode} 
\usepackage{bm}
\usepackage{graphbox}
\usepackage[colorlinks,linkcolor=black,anchorcolor=black,
citecolor=black,urlcolor=black]{hyperref}

\usepackage{stfloats}
\usepackage{siunitx}
\DeclareSIUnit\bps{bps}
\DeclareSIUnit\Torr{Torr}
\DeclareSIUnit\torr{Torr}
\DeclareSIUnit\sample{Sa}
\newcommand*{\circled}[1]{\lower.7ex\hbox{\tikz\draw (0pt, 0pt)%
  circle (.5em) node {\makebox[1em][c]{\small #1}};}}

\ifCLASSINFOpdf

\else

\fi

\graphicspath{{figures/}}

\begin{document}

\title{DSS-o-SAGE: Direction-Scan Sounding-Oriented SAGE Algorithm for Channel Parameter Estimation in mmWave and THz Bands}
\author{Yuanbo~Li, Chong~Han,~\IEEEmembership{Senior~Member,~IEEE}, Yi Chen, Ziming Yu, and Xuefeng Yin,~\IEEEmembership{Senior~Member,~IEEE}
\thanks{
Yuanbo Li, and Chong Han are with the Terahertz Wireless Communications (TWC) Laboratory, Shanghai Jiao Tong University, Shanghai, China (e-mail: \{yuanbo.li,chong.han\}@sjtu.edu.cn).

Yi Chen and Ziming Yu are with Huawei Technologies Co., Ltd, China (e-mail:\{chenyi171,yuziming\}@huawei.com).

Xuefeng Yin are with the College of Electronics and Information Engineering, Tongji University, Shanghai, China (e-mail: yinxuefeng@tongji.edu.cn).
}
}

	{}
	\maketitle
	\thispagestyle{empty}
\begin{abstract}
Investigation of millimeter (mmWave) and Terahertz (THz) channels relies on channel measurements and estimation of multi-path component (MPC) parameters. As a common measurement technique in the mmWave and THz bands, direction-scan sounding (DSS) resolves angular information and increases the measurable distance. Through mechanical rotation, the DSS creates a virtual multi-antenna sounding system, which however incurs signal phase instability and large data sizes, which are not fully considered in existing estimation algorithms and thus make them ineffective. To tackle this research gap, in this paper, a DSS-oriented space-alternating generalized expectation-maximization (DSS-o-SAGE) algorithm is proposed for channel parameter estimation in mmWave and THz bands. To appropriately capture the measured data in mmWave and THz DSS, the phase instability is modeled by the scanning-direction-dependent signal phases. Furthermore, based on the signal model, the DSS-o-SAGE algorithm is developed, which not only addresses the problems brought by phase instability, but also achieves ultra-low computational complexity by exploiting the narrow antenna beam property of DSS. 
Simulations in synthetic channels are conducted to demonstrate the efficacy of the proposed algorithm and explore the applicable region of the far-field approximation in DSS-o-SAGE. Last but not least, the proposed DSS-o-SAGE algorithm is applied in real measurements in an indoor corridor scenario at 300~GHz. Compared with results using the baseline noise-elimination method, the channel is characterized more correctly and reasonably based on the DSS-o-SAGE.
\end{abstract}
\section{Introduction}
\par \IEEEPARstart{D}{uring} the last several decades, the communication community has witnessed a giant leap of communication technologies, from the first generation to the fifth generation mobile communication systems (5G). To support the large number of intelligent devices and applications, such as metaverse, autonomous driving, etc., it is believed that the data rate in the sixth generation mobile communication system (6G) will need to exceed hundreds of gigabits per second and even Terabits per second to undertake the explosively grown data traffic~\cite{chen2021terahertz}. As a result, the millimeter-wave (mmWave) and Terahertz (THz) bands, ranging from \SI{0.03}{THz} to \SI{10}{THz}, are envisioned as a key technology to enable such high data rate~\cite{xiao2017mmwave,akyildiz2018combating,rappaport2019wireless}. With abundant spectrum resources and ultra-large bandwidth (more than tens of GHz), the use of mmWave and THz bands can address the spectrum scarcity and capacity limitations of current wireless systems.
\par However, challenges remain to achieve reliable mmWave and THz communications, among which a fundamental one lies in the channel modeling in mmWave and THz bands. To thoroughly investigate propagation phenomena in mmWave and THz bands, channel measurement campaigns are the most convincing and thus popular way to obtain realistic data. To obtain angular information on wireless channels, channel measurements can be conducted using multi-input multi-output (MIMO) sounding or direction-scan sounding (DSS). Still due to immature antenna array techniques, mmWave and THz channel measurement campaigns are usually conducted based on DSS, which involves installing directional antennas on rotators and mechanically changing the pointing directions of transmitters (Tx) and receivers (Rx) to scan the spatial domain, creating a \textit{virtual multi-antenna sounder system}. Many research groups have utilized this method to measure mmWave and THz channels, in both indoor scenarios~\cite{Erden202028,Rappaport2015Wideband,Xing2021mmwave,ju2021millimeter,yi2021channel,cheng2019characterization,priebe2011channel,he2021channel,li2022channel} and outdoor scenarios~\cite{abbasi2023thz,eckhardt2021channel,Lyu2023Measurement}, as summarized in~\cite{han2022terahertz}. 
\par Channel analysis and modeling rely on accurate parameter estimation of multi-path components (MPCs), which can be achieved by using high-resolution parameter estimation (HRPE) algorithms. Generally speaking, HRPE algorithms can be divided into three categories, including spectral-based methods~\cite{capon1969high,bartlett1948smoothing}, parametric subspace-based methods~\cite{Zhang2017Channel,Guo2017Millimeter}, and maximum-likelihood (ML) based methods~\cite{thoma2004rimax,moon1996expectation}. Among the three kinds, the ML-based methods, such as Richter’s ML estimation (RiMAX)~\cite{thoma2004rimax} and expectation maximization (EM) algorithms~\cite{moon1996expectation}, possess high resolution but also high complexity. Nonetheless, as an extension of the EM algorithm, the space-alternating generalized expectation-maximization (SAGE) algorithm significantly reduces the complexity by iteratively estimating channel parameters one path by another~\cite{Fessler1994Space,fleury1999channel}, which has been the most widely used and extended in the last two decades~\cite{Shutin2011sparse,ou2016sage,yin2016performance,yin2017scatter,jiang2022arma,hong2023joint,Zhou2023Novel}. {For instance, by exploring the Bayesian methods, \textit{D. Shutin et al.} proposed a sparse variational Bayesian SAGE (VB-SAGE) algorithm, to address the inaccuracy of SAGE under influences of diffuse scattering and possible calibration errors~\cite{Shutin2011sparse}. \textit{L. Ouyang et al.} proposed a signal eigenvector-based-SAGE (SEV-SAGE) algorithm for channel parameter estimation in low signal-to-noise ratio (SNR) scenarios~\cite{ou2016sage}. Moreover, \textit{X. Yin et al.} extended the applicable range of the SAGE algorithm from the far-field to the near-field, which is then used for scatter localization~\cite{yin2017scatter}. In their subsequent work, a geometry-aided SAGE (GA-SAGE) algorithm was proposed~\cite{hong2023joint}. By assuming the scatter locations on certain geometries, such as ellipses for single-bounce paths, the computational complexity of near-field channel parameter estimation is greatly reduced. Furthermore, \textit{Z. Zhou et al.} proposed a novel SAGE algorithm for parameter estimations of wideband spatial non-stationary wireless channels with antenna polarization (SAGE-WSNSAP)~\cite{Zhou2023Novel}.} 

\par {However, to the best of the authors' knowledge, none of the existing SAGE extensions are dedicated to mmWave and THz DSS, for which they cannot be effectively applied due to the phase instability and large data sizes in mmWave and THz DSS. First, existing SAGE algorithms make the assumption of stable signal phases across the array, which is violated in mmWave and THz DSS.} On one hand, the radiation pattern of the horn antenna results in different signal phases in different scanning directions. Although this term is deterministic, due to system non-idealities and uncertainties, e.g. antenna phase center positioning error, cable movement, etc., accurate phase pattern measurement is difficult in mmWave and THz bands~\cite{li2023Antenna}. On the other hand, during the DSS process, stochastic phase variations will occur due to instabilities of the measurement system, such as RF system instabilities, cable bending effects, random vibration of the rotators, etc., which are hard to predict and can only be compensated for with additional hardware costs~\cite{Parssinen20216gwhitepaper,Lyu2021Design,Lyu2023Enabling,Lyu2023Virtual}. {Second, although several works proposed some methods to reduce the computational complexity, such as the coarse-to-fine estimation process and simplified signal models~\cite{Cai2020Trajectory,Zhou2023Novel,hong2023joint}, it still requires large computational and time resources for channel parameter estimation due to the large data size in mmWave and THz DSS.} Since the mmWave and THz DSS are usually measured with large bandwidth, the measured CIRs usually include hundreds or thousands of sample points, for which the total time and computation resources consumption for channel parameter estimation is very high. Therefore, further simplifications need to be considered to reduce the computational complexity.
\par {Apart from the above-mentioned challenges, research gaps also exist in the investigation of near-field effects and the performance evaluation of the noise-elimination method in mmWave and THz DSS. First, the effects of near-field are not fully investigated for channel parameter estimations in mmWave and THz DSS.} Judging by the widely used Rayleigh distance boundary between far- and near-fields, calculated as $d=2L_a^2/\lambda$, with $L_a$ denoting the antenna aperture and $\lambda$ being the wavelength, the Tx/Rx are usually in the near-field due to the large aperture of the virtual spherical array (VSA) in DSS and small wavelength in mmWave and THz bands. However, since the antennas used in mmWave and THz DSS are directional, the near-field effects may not be as severe as expected. The impact of near-field on SAGE algorithm performance in mmWave and THz DSS estimation remains an open problem. {Second, without effective HRPE algorithms, channel measurement results in mmWave and THz DSS are usually processed with noise-elimination methods, whose performance is not carefully testified.} Current channel measurements in mmWave and THz bands use signal processing methods to estimate MPC parameters and extract channel characteristics, such as noise-elimination methods~\cite{Xing2021mmwave,ju2021millimeter,yi2021channel,cheng2019characterization,priebe2011channel,he2021channel,li2022channel,abbasi2023thz,eckhardt2021channel} and beamforming methods~\cite{li2022Virtual,li2023Antenna,yuan2023Millimeter}. Though the beamforming methods are much more effective than noise-elimination methods, they require dense scanning in the spatial domain and thus large time consumption, for which the noise-elimination method is preferred and widely used in the literature. However, despite its ease of implementation, the noise-elimination method lacks sufficient accuracy and fails to decouple the antenna effects in DSS. As a result, it may produce inaccurate channel characteristics and even lead to erroneous conclusions. Therefore, it is necessary to carefully and thoroughly analyze how the coupled antenna effects in the method affect the channel characterization results.
\par {To address the aforementioned challenges and research gaps, in this paper, we propose a DSS-oriented SAGE (DSS-o-SAGE) algorithm for channel parameter estimation in mmWave and THz DSS. The phase instability is characterized by scanning-direction-dependent phases in the signal model, which is further validated with real measurements. In addition, the DSS-o-SAGE algorithm is derived, where low computational complexity is achieved by exploring the narrow antenna beam in mmWave and THz DSS and using a coarse-to-fine search process with partial data. Furthermore, to demonstrate the efficacy of the proposed algorithm, its performance is compared with existing SAGE algorithms through simulations as well as realistic measurements. The results show that with special considerations for mmWave and THz DSS, the DSS-o-SAGE algorithm has higher estimation accuracy and lower computational complexity compared to existing methods. Extensive simulations are performed to model the scatter distance boundary, beyond which the far-field simplification can be used in DSS-o-SAGE.} By using the proposed DSS-o-SAGE algorithm, the channel characterization in mmWave and THz bands is more accurate and reasonable. Distinctive contributions of our work are summarized as follows.
\begin{itemize}
    \item {\textbf{We introduce scanning-direction-dependent phases to model the phase instability in mmWave and THz DSS.} To validate the existence of phase instability, channel measurements are conducted in a small chamber. The results show that the signal phases are unstable during DSS.}
    \item \textbf{We propose a DSS-o-SAGE algorithm, which is useful for channel parameter estimation in mmWave and THz DSS.} {The maximum likelihood estimator (MLE) is derived taking into account the scanning-direction-dependent phases.} To reduce computational complexity, a coarse-to-fine process is used to decompose the high-dimensional estimation problem. In addition, utilizing the narrow antenna beam property, only partial data near the coarse estimations is used, which significantly reduces the time and computation resource consumption.
    
    \item \textbf{We evaluate the performance of the DSS-o-SAGE algorithm in synthetic channels.} {Comparisons are made with existing SAGE algorithms to demonstrate the superiority of the DSS-o-SAGE algorithm in terms of computational complexity and estimation accuracy. Moreover, the applicable region of near-and far-field assumptions, when using the DSS-o-SAGE algorithm, is modeled.}
    \item \textbf{We apply the proposed DSS-o-SAGE algorithm in real measurement campaigns in an indoor corridor scenario at 300~GHz}. Based on the estimated MPC parameters, propagation analysis and channel characteristics are elaborated and compared with those based on the noise-elimination method. By using the DSS-o-SAGE algorithm, fake MPCs caused by the influence of antenna radiation patterns are eliminated, for which different MPCs can be more easily distinguished and channel characterizations differ from those obtained by using noise-elimination methods.
\end{itemize}
\par The remainder of the paper is organized as follows. The signal model with phase instability is introduced and validated in Sec.~\ref{sec:signalmodel}. Section~\ref{sec:nnsage} provides a detailed derivation and explanation of the DSS-o-SAGE algorithm. Furthermore, in Sec.~\ref{sec:com}, the performance of the DSS-o-SAGE algorithm and influences of near-field are studied based on simulations. Moreover, real measured data in an indoor corridor scenario is used to compare the DSS-o-SAGE algorithm with existing methods in Sec.~\ref{sec:char}. Finally, Sec.~\ref{sec:conclude} concludes the paper.

\section{Signal Model and Experimental Validation}
\label{sec:signalmodel}
\par In this section, the signal model involving the phase instability of the mmWave and THz DSS is introduced. Moreover, experiments are conducted for validation.
\subsection{Signal Model}
\label{sec:signalmodel1}
\begin{figure*}
    \centering
    \includegraphics[width=1.7\columnwidth]{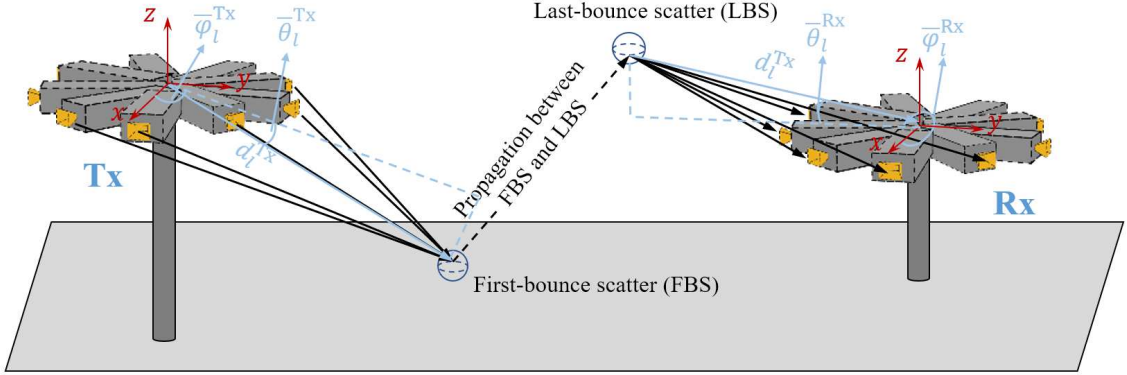}
    \caption{Schematic diagram for mmWave and THz DSS.}
    \label{fig:signalmodel}
    \vspace{-0.5cm}
\end{figure*}
\par In mmWave and THz DSS, the Tx/Rx radio frequency (RF) front ends are installed on gimbals/rotators, by which the pointing directions can be mechanically changed to resolve the spatial information of MPCs, as imaged in Fig.~\ref{fig:signalmodel}. The diameter of the VSA formed by the rotation of the RF front end is typically tens of centimeters, i.e., hundreds of wavelengths. As a result, evaluated using the Rayleigh distance, the Tx and Rx are usually in the near-field and the MPC parameters when Rx points towards different scanning directions should be calculated using the spherical wavefront.
\par During the propagation of MPCs, it is considered that the first bounce scatter (FBS) and last bounce scatter (LBS) are acting as point sources radiating to VSA of Tx/Rx~\cite{yin2017scatter,7744514}, as shown in Fig.~\ref{fig:signalmodel}.
Specifically, considering the $(n_t)^\text{th}$ scanning direction of Tx and $(n_r)^\text{th}$ scanning direction of Rx, the delay, direction-of-departure (DoD) and direction-of-arrival (DoA) of the $l^\text{th}$ MPC can be expressed as
\begin{align}
    \tau_{l,n_t,n_r}&=\tau_l+\frac{||\bm{r}^{\text{FBS}}-\bm{r}^{\text{Tx}}_{n_t}||-d^\text{Tx}_{l}}{c}\nonumber\\
    &~~~~~~+\frac{||\bm{r}^{\text{LBS}}-\bm{r}^{\text{Rx}}_{n_r}||-d^\text{Rx}_{l}}{c},
    \label{eq:losdelay}\\
    \bm{\Omega}^\text{Tx}_{l,n_t}&=\frac{\bm{r}^{\text{FBS}}-\bm{r}^{\text{Tx}}_{n_t}}{||\bm{r}^{\text{FBS}}-\bm{r}^{\text{Tx}}_{n_t}||},\\
    \bm{\Omega}^\text{Rx}_{l,n_r}&=\frac{\bm{r}^{\text{LBS}}-\bm{r}^{\text{Rx}}_{n_r}}{||\bm{r}^{\text{LBS}}-\bm{r}^{\text{Rx}}_{n_r}||},
    \label{eq:losdoa}
\end{align}
where $c$ is the speed of light in free space. $\tau_l$ is the propagation delay from the reference Tx position to the reference Rx position set as the center of the Tx/Rx VSA, as indicated in Fig.~\ref{fig:signalmodel}. Moreover, $\bm{r}^{\text{Tx}}_{n_t}$ and $\bm{r}^{\text{Rx}}_{n_r}$ are antenna locations of Tx in the $(n_t)^\text{th}$ Tx scanning direction and that of Rx in the $(n_r)^\text{th}$ Rx scanning direction, respectively. Furthermore, the locations of FBS $\bm{r}^{\text{FBS}}$ and LBS $\bm{r}^{\text{LBS}}$ are expressed as
\begin{equation}
    \begin{split}
    \bm{r}^{\text{FBS}}=\overline{\bm{r}}^\text{Tx}+d^\text{Tx}_l\overline{\bm{\Omega}}^\text{Tx}_l,\\
    \bm{r}^{\text{LBS}}=\overline{\bm{r}}^\text{Rx}+d^\text{Rx}_l\overline{\bm{\Omega}}^\text{Rx}_l,
    \label{eq:lbspos}
\end{split}
\end{equation}
where $\overline{\bm{r}}^\text{Tx}$ and $\overline{\bm{r}}^\text{Rx}$ are the locations of reference Tx/Rx positions. Moreover, $\overline{\bm{\Omega}}^\text{Tx}_l$ and $\overline{\bm{\Omega}}^\text{Rx}_l$ are the DoD and DoA observed in the reference Tx/Rx position, expressed as
\begin{equation}
    \overline{\bm{\Omega}}^a_l=[\cos(\varphi^a_{l})\cos(\theta^a_{l}),\sin(\varphi^a_{l})\cos(\theta^a_{l}),\sin(\theta^a_{l})],
\end{equation}
where $a$ could be either "Tx" or "Rx", representing DoD or DoA. Additionally, $\varphi^a_l$ and $\theta^a_l$ are the azimuth angle and elevation angle of the $l^\text{th}$ path observed in the reference Tx/Rx position, respectively.
\par Note that the above equations are derived considering a spherical wavefront (SWF), which always hold for both the far-field and near-field. {When the influences of the near-field are not significant, as will be discussed in Sec.~\ref{sec:nearfield}, far-field approximations (FFA) can be made by simply letting the Rx-LBS distance and Tx-FBS distance approach infinity.}
\par To cover the spatial area of interest, we assume that in total $N_t$ and $N_r$ directions are scanned at the Tx and Rx sides. Even though the CIRs may be measured using different kinds of equipment, such as time-domain based and frequency-domain based systems~\cite{han2022terahertz}, the CIRs can be expressed in a general form, as derived in Appendix~\ref{appen:cir}. Specifically, for the $(n_r)^\text{th}$ pointing direction at Rx, the CIR is measured as
\begin{equation}
\begin{split}
    h_{n_t, n_r}[i]=\sum_{l=1}^L\alpha_l&c^\text{Tx}_{n_t}(\bm{\Omega}^\text{Tx}_{l,n_t})c^\text{Rx}_{n_r}(\bm{\Omega}^\text{Rx}_{l,n_r})\text{e}^{j\phi_{l,n_t,n_r}}\cdot\\
    &R_{\tau_{l,n_t,n_r}}[i]+w_{n_t,n_r}[i],
    \label{eq:dcir}
\end{split}
\end{equation}
where $i$ is the index of temporal samples and $L$ is the number of MPCs. Besides, $\alpha_l$, $\tau_{l,n_t,n_r}$, $\bm{\Omega}^\text{Tx}_{l,n_t}$,$\bm{\Omega}^\text{Rx}_{l,n_r}$, $\phi_{l,n_t,n_r}$ denote the real-valued path gain, delay, DoD, DoA, and phase of the $l^\text{th}$ MPC, respectively. $R_{\tau_{l,n_t,n_r}}[i]$ has a similar shape with Dirac functions and is derived in Appendix~\ref{appen:cir}. $w_{n_t,n_r}[i]$ represents the noise components. Furthermore,$c^\text{Tx}_{n_t}(\cdot)$ and $c^\text{Rx}_{n_r}(\cdot)$ stand for the real-valued radiation pattern of the Tx/Rx horn antenna, respectively. 
\par {Note that the one major difference between the signal model in this work and those in existing works in~\cite{yin2017scatter,hong2023joint,7744514} is the scanning-direction-dependent signal phase $\phi_{l,n_t,n_r}$. Due to the aforementioned effects of antenna radiation pattern and system instabilities, we consider the signal phases in different scanning directions as independent parameters, rather than a consistent parameter across the array. As will be shown in Sec. IV, such a consideration makes our algorithm robust under the phase instability in mmWave and THz DSS.}
\par By combining the CIRs in different scanning directions together, we can obtain that
\begin{equation}
\begin{split}
    \bm{h}&=\sum_{l=1}^L\bm{s}(\bm{\theta}_l)+\bm{w}\\
    &=\bm{s}(\bm{\Theta})+\bm{w},
    \label{eq:cirtensor}
\end{split}
\end{equation}
where $\bm{h}$ and $\bm{w}$ are $N_t\times N_r\times I$ tensors, containing the CIR data and noise samples, respectively. Moreover, $\bm{\theta}_l=[\alpha_l,\tau_l,\varphi^\text{Tx}_l,\theta^\text{Tx}_l,\varphi^\text{Rx}_l,\theta^\text{Rx}_l,d^\text{Tx}_l,d^\text{Rx}_l,\phi_{l,1,1},...,\phi_{l,N_t,N_r}]$ includes the parameters of the $l^\text{th}$ path. Moreover, $\bm{\Theta}=[\bm{\theta}_1,...,\bm{\theta}_L]$ includes parameters of all MPCs. Furthermore, $\bm{s}(\bm{\theta}_l)$ is a $N_t\times N_r\times I$ tensor representing the signal component of the $l^\text{th}$ MPC, whose $(n_t,n_r,i)^\text{th}$ element is the $l^\text{th}$ item of the summation on the right side of~\eqref{eq:dcir}.
\subsection{Experimental Validation of the Phase Instability}
\label{sec:pattern}
\begin{figure}
    \centering
    \subfloat[LoS experiment.]{
    \includegraphics[width=0.24\textwidth]{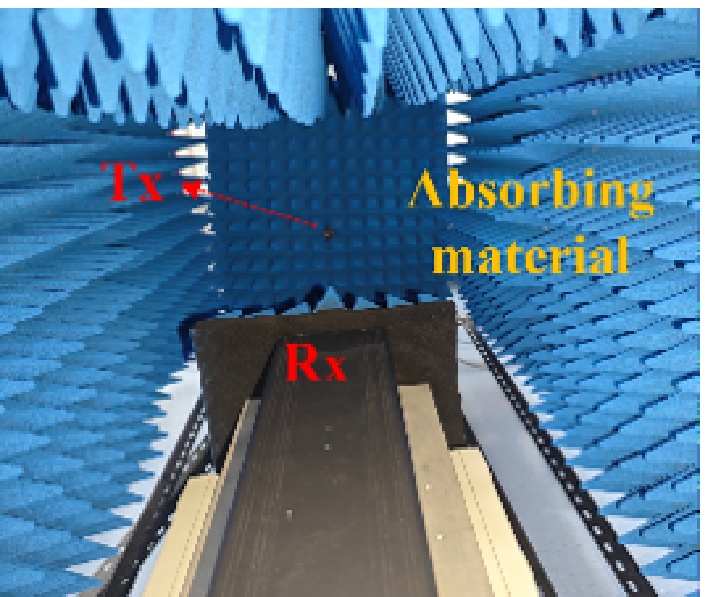}
    }
    \subfloat[Reflection experiment.]{
    \includegraphics[width=0.24\textwidth]{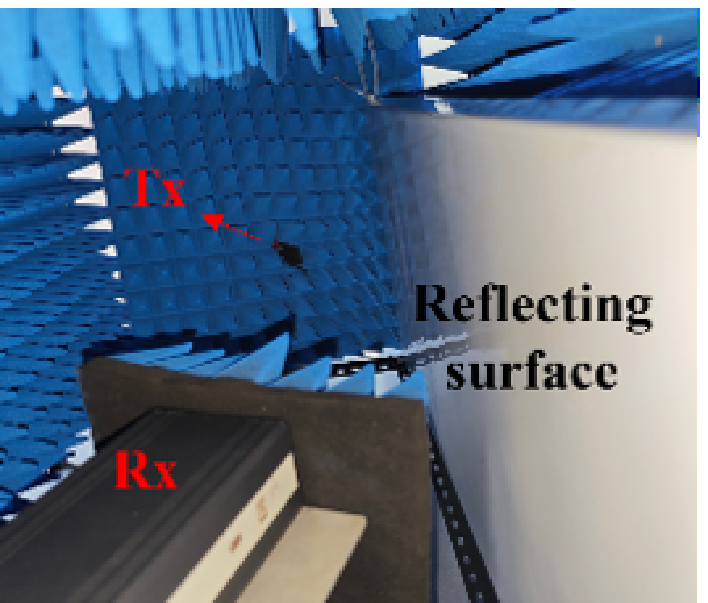}
    }
    \caption{Picture of the conducted experiment.}
    \label{fig:exp}
    \vspace{-0.5cm}
\end{figure}
To validate the aforementioned phase instability in different scanning directions, experiments are conducted using a vector network analyzer (VNA)-based channel sounder in a small chamber, as shown in Fig.~\ref{fig:exp}. Two experiments are conducted, including a LoS experiment and a reflection experiment. For the LoS experiment, the Tx and Rx are surrounded by absorbing materials, to avoid interference of other MPCs. In contrast, a metal surface is added in the reflection experiment. {Note that since the edges of the reflecting surface are not sharp and both Tx and Rx are on the same side of the surface, the diffracted path by the reflecting surface is omitted here.} In both experiments, Tx and Rx are separated by \SI{0.6}{m}. Besides, the azimuth angles of arrival (AoA) of the LoS path and the reflected path are $0^\circ$ and $28^\circ$, respectively, while the elevation angles of arrival (EoA) are $0^\circ$ for both LoS and reflection experiments. {Moreover, the measured frequency band ranges from \SIrange{306}{321}{GHz}, with \SI{15}{GHz} bandwidth, achieving a time resolution of \SI{0.067}{ns}}. Both Tx and Rx are equipped with horn antennas, whose gain is \SI{25}{dBi} and half-power beam width (HPBW) is $8^\circ$. For both experiments, the Tx remains static, while the Rx scans the spatial domain, from $-20^\circ$ to $20^\circ$ in the elevation plane and from $-20^\circ$ to $20^\circ$ in the LoS experiment, $10^\circ$ to $50^\circ$ in the reflection experiment, in the azimuth plane, respectively. The spatial scanning step is $2.5^\circ$. {Note that since the experimental results here are only used to observe the phase instability, such a $2.5^\circ$ scanning step is sufficient.}
\begin{figure}
    \centering
    \includegraphics[width=0.6\columnwidth]{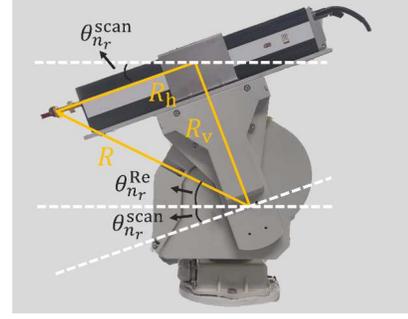}
    \caption{The rotator used for channel measurements.}
    \label{fig:rotator}
    \vspace{-0.5cm}
\end{figure}
\par The rotator used to change the Rx scanning direction is shown in Fig.~\ref{fig:rotator}. Taking the center of the rotator as the coordinate origin, the position of the horn antenna in the $(n_r)^\text{th}$ scanning direction can be expressed as
\begin{equation}
\bm{r}^\text{Rx}_{n_r}=R[\cos(\varphi^\text{Re}_{n_r})\cos(\theta^\text{Re}_{n_r}),\sin(\varphi^\text{Re}_{n_r})\cos(\theta^\text{Re}_{n_r}),\sin(\theta^\text{Re}_{n_r})],
\end{equation}
where $\varphi^\text{Re}_{n_r}$ and $\theta^\text{Re}_{p,n_r}$ are the azimuth and elevation angles of the horn antenna relative to the center of the rotator, respectively. Moreover, $R$ is the rotation radius. Based on the geometric relations shown in Fig.~\ref{fig:rotator}, these parameters are
\begin{align}
    \varphi^\text{Re}_{n_r}&=\varphi^{\text{scan}}_{n_r},\\
    \theta^\text{Re}_{n_r}&=\theta^{\text{scan}}_{n_r}+\arctan\left(\frac{R_\text{v}}{R_\text{h}}\right),\\
    R&=\sqrt{R_\text{v}^2+R_\text{h}^2},
\end{align}
where $\varphi^{\text{scan}}_{n_r}$ and $\theta^{\text{scan}}_{n_r}$ are the azimuth and elevation angles of the pointing direction of the Rx horn antenna, respectively. Additionally, $R_\text{h}$ and $R_\text{v}$ are the horizontal and vertical radius of the rotator, as shown in Fig.~\ref{fig:rotator}. In this work, the horizontal and vertical radii are \SI{0.23}{m} and \SI{0.18}{m}, resulting in a VSA radius of \SI{0.29}{m} and a Rayleigh distance of \SI{672.8}{m} at \SI{300}{GHz}.
\par The signal phase of the LoS/reflected path can be estimated using a simple maximum likelihood estimator (MLE), as
\begin{align}
    \hat{\phi}_{n_r}=\text{Angle}(\sum_{i=1}^Ih_{n_r}[i]R_{\hat{\tau}_{n_r}}[i]),
    \label{eq:mlesignal}
\end{align}
where $I$ is the number of temporal samples. Moreover, $\text{Angle}(\cdot)$ calculates the phase of complex numbers. $\hat{\tau}_{n_r}$ is the estimated delay in the $n_r^\text{th}$ scanning direction, as
\begin{equation}
    \hat{\tau}_{n_r}=\underset{\tau}{{\arg\max}}\frac{\left|\sum_{i=1}^Ih_{n_r}[i]R_\tau[i]\right|^2}{\sum_{i=1}\left|R_\tau[i]\right|^2}.
\end{equation}

The derivation of this MLE is omitted here, since it is a simplified case of the MLE derived in Sec.~\ref{sec:mstep}.
\begin{figure}
    \centering
    \subfloat[LoS experiment.]{
    \includegraphics[width=0.24\textwidth]{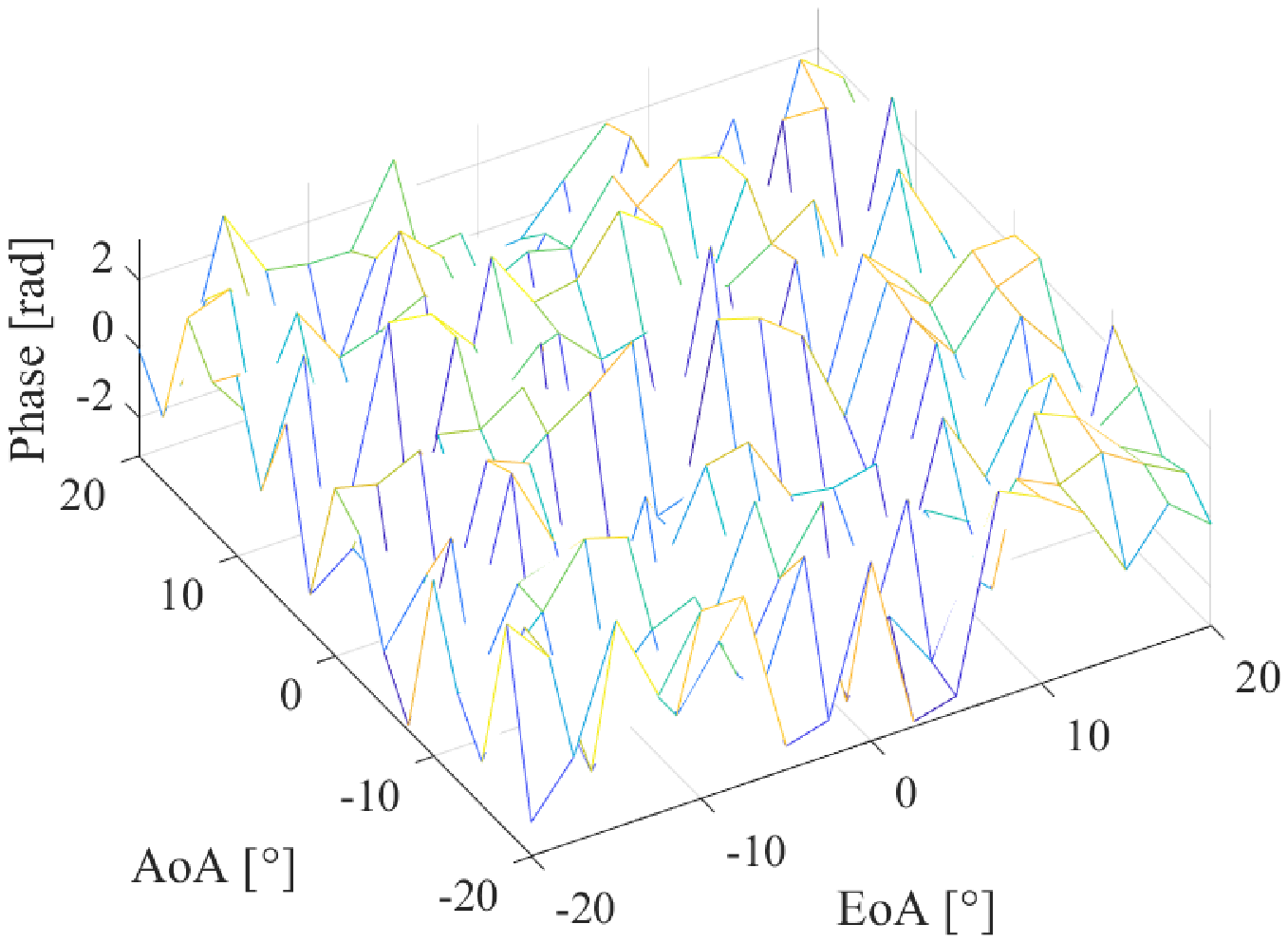}
    }
    \subfloat[Reflection experiment.]{
    \includegraphics[width=0.24\textwidth]{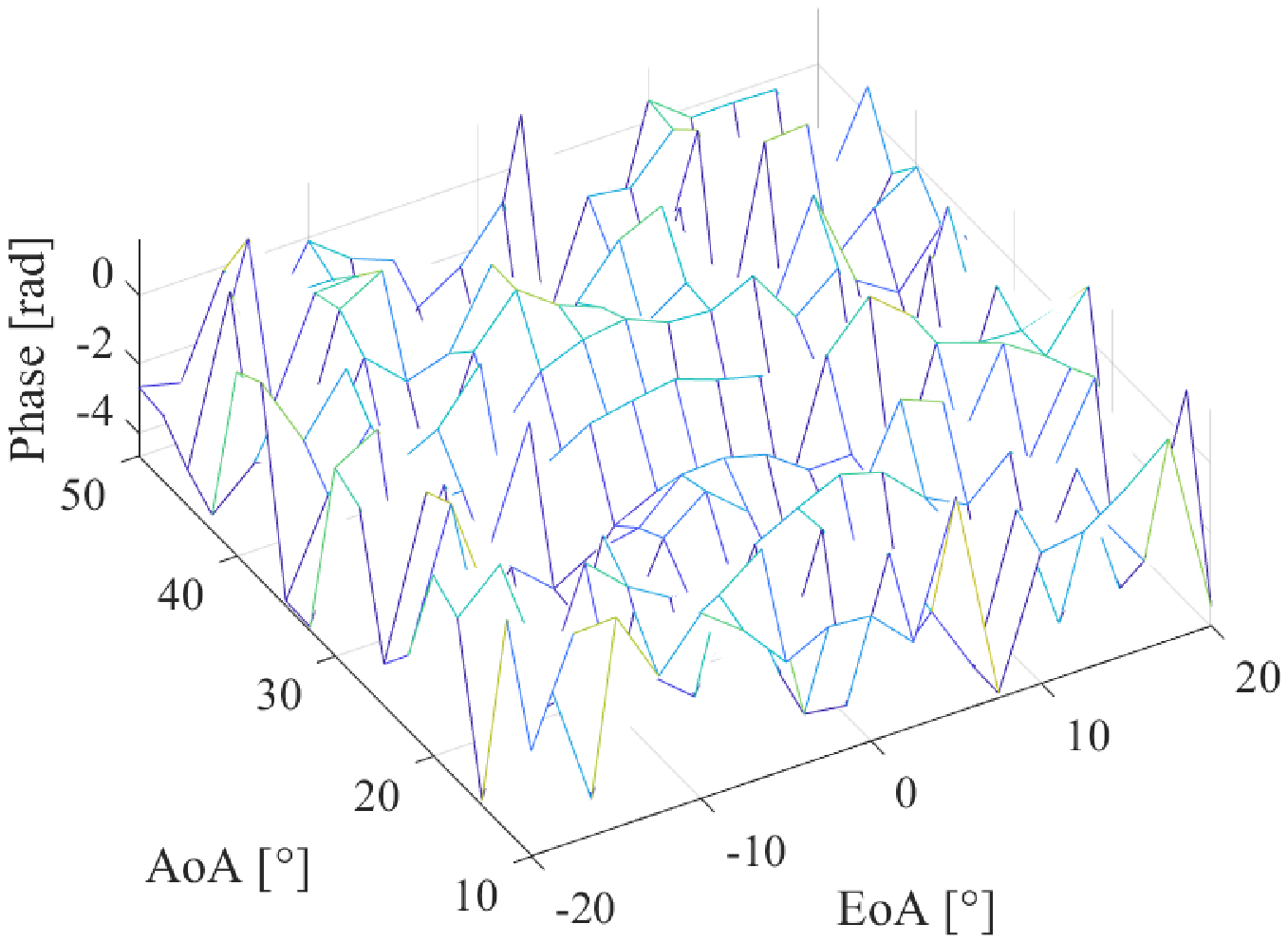}
    }
    \caption{Signal phase in different scanning directions.}
    \label{fig:phase}
    \vspace{-0.5cm}
\end{figure}
The estimated signal phases using~\eqref{eq:mlesignal} are shown in Fig.~\ref{fig:phase}. It can be observed that the signal phase is not stable for different scanning directions. For those scanning angles far from the DoA ([$0^\circ,0^\circ$] for the LoS experiment and [$0^\circ,28^\circ$] for the reflection experiment), the phase instability is significant. For those signals close to the DoA of MPCs, the signal phase also vibrates, due to reasons like antenna radiation pattern, cable movement, among others. {To regenerate the scanning-direction-dependent phases in simulations in Sec.~\ref{sec:simulation}, the signal phases are normalized to that in the strongest direction, which are further fitted with normal distributions. The mean values are 0.13 rad and 0.07 rad for LoS and reflection experiments, respectively, and the standard deviations are 1.83 and 1.88 for LoS and reflection experiments, respectively.}
\section{DSS-o-SAGE Algorithm for Channel Parameter Estimation}
\label{sec:nnsage}
\par With the signal model in~\eqref{eq:cirtensor}, assuming that the noise samples are independent random variables following complex Gaussian distributions, the log-likelihood function of $\bm{\Theta}$ given a measurement data $\bm{h}$ can be easily derived as~\cite{fleury1999channel}
\begin{equation}
    \Lambda(\bm{\Theta};\bm{h})\propto -||\text{vec}[\bm{h}]-\text{vec}[\bm{s}(\bm{\Theta})]||^2,
    \label{eq:likeliall}
\end{equation}
where $\text{vec}[\cdot]$ represents the vectorization operation that converts tensors to column vectors. 
\begin{figure*}[!tbp]
    \centering
    \includegraphics[width=1.9\columnwidth]{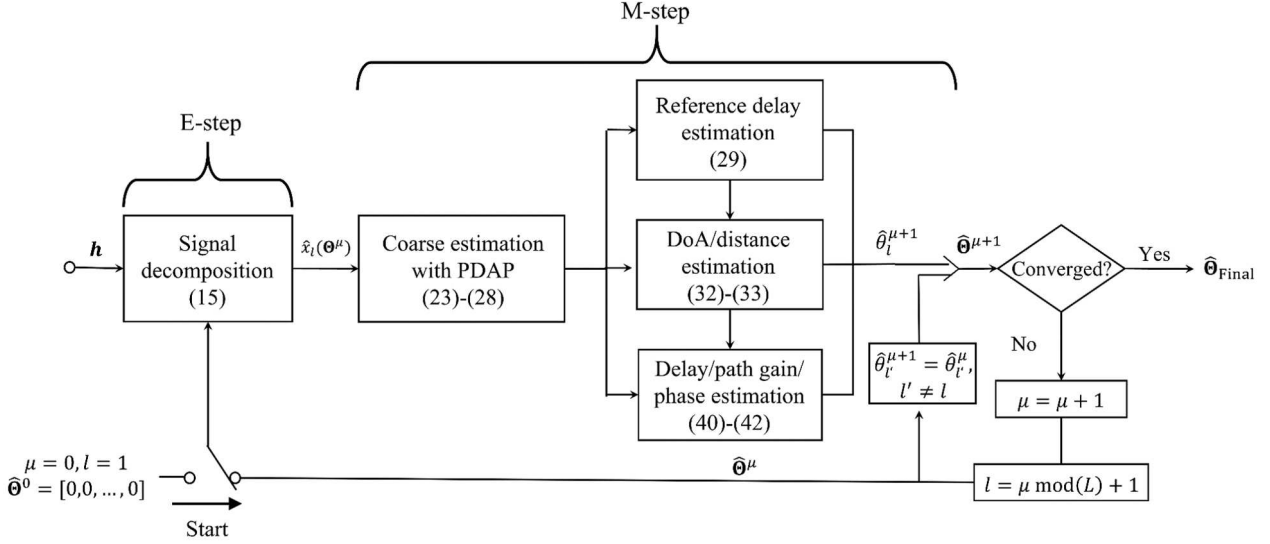}
    \caption{The flowchart of the DSS-o-SAGE algorithm.}
    \label{fig:sage}
    \vspace{-0.5cm}
\end{figure*}
\par The MLE of $\bm{\Theta}$, namely $\hat{\bm{\Theta}}_{\text{ML}}$, is given by $\hat{\bm{\Theta}}_{\text{ML}}=\arg\max_{\bm{\Theta}}~\Lambda(\bm{\Theta};\bm{h})$. Without closed-form solutions, the MLE must be calculated numerically. Due to the ultra-large dimensions of the parameter space, directly solving the MLE is computationally impossible. To overcome this, the DSS-o-SAGE algorithm is proposed, whose flowchart is shown in Fig.~\ref{fig:sage}. The DSS-o-SAGE algorithm iteratively maximizes the likelihood function, where each iteration consists of an expectation (E)-step and a maximization (M)-step. In the E-step, the signal component of a certain MPC is estimated by subtracting effects of other MPCs from the measured data. In the M-step, the parameters of the concerned MPC are updated by maximizing the log-likelihood function given the signal component estimation. The MPC parameters are updated path by path, until the overall likelihood function converges.
\subsection{E-step}
\par In the $\mu^\text{th}$ iteration, the component of the $l^\text{th}$ path $\bm{s}(\bm{\theta}_l)$ is estimated as the conditional expectation given the measured data $\bm{h}$ and the MPC parameter estimations $\bm{\Theta}^\mu$, as
\begin{equation}
\begin{split}
    \hat{\bm{x}}_l(\hat{\bm{\Theta}}^{\mu})&=\bm{E}_{\bm{\Theta}^\mu}[\bm{s}(\bm{\theta}_l)|\bm{h}]
    \\&=\bm{h}-\sum_{l'=1,l'\neq l}^{L}\bm{s}(\hat{\bm{\theta}}^{\mu}_{l'})
    \\&=\bm{s}(\bm{\theta}_l)+\bm{w}',
    \label{eq:signaleach}
\end{split}
\end{equation}
where the term $\bm{w}'$ includes not only the noise $\bm{w}$ but also residual errors due to inaccurate estimations of $(l')^\text{th}$ MPC.
\par It can be noticed that~\eqref{eq:signaleach} is in a similar form as~\eqref{eq:cirtensor}, for which the log-likelihood function of $\bm{\theta}_l$ given the signal component estimation $\hat{\bm{x}}_l(\hat{\bm{\Theta}}^{\mu})$ is expressed as,
\begin{equation}
    \Lambda(\bm{\theta}_l;\hat{\bm{x}}_l(\hat{\bm{\Theta}}^{\mu}))\propto -||\text{vec}[\hat{\bm{x}}_l(\hat{\bm{\Theta}}^{\mu}])-\text{vec}[\bm{s}(\bm{\theta}_l)]||^2.
    \label{eq:likelieach}
\end{equation}
\par Therefore, with the likelihood function, the estimations of the MPC parameters of the $l^\text{th}$ path can be solved using the MLE, which is done in the following M-step.
\subsection{M-step}
\label{sec:mstep}
\par The likelihood function in~\eqref{eq:likelieach} can be simplified as
\begin{equation}
\begin{split}
    \Lambda\propto\sum_{n_t,n_r,i}&2\mathcal{R}\left\{[\bm{s}(\bm{\theta}_l)\odot(\hat{\bm{x}}_{l}(\hat{\bm{\Theta}}^{\mu}))^*]_{n_t,n_r,i}\right\}\\
    &-\sum_{n_t,n_r,i}\left|[\bm{s}(\bm{\theta}_l)]_{n_t,n_r,i}\right|^2,
    \label{eq:likeli_nosimplify}
\end{split}
\end{equation}
where $\bm{A}\odot\bm{B}$ denotes the Hadamard product  between two matrices of the same size, and $(\cdot)^*$ is the conjugate of complex numbers. Moreover, $[\cdot]_{n_1,n_2,n_3}$ stands for the corresponding element of the matrix corresponding to indices $n_1,n_2,n_3$. Additionally, the signal component $[\bm{s}(\bm{\theta}_l)]_{n_t,n_r,i}$ is equal to $\alpha_l\text{e}^{j\phi_{l,n_t,n_r}}c^\text{Tx}_{n_t}(\bm{\Omega}^\text{Tx}_{l,n_t})c^\text{Rx}_{n_r}(\bm{\Omega}^\text{Rx}_{l,n_r})R_{\tau_{l,n_t,n_r}}[i]$.
\par Substituting the expressions of $[\bm{s}(\bm{\theta}_l)]_{n_t,n_r,i}$ into~\eqref{eq:likeli_nosimplify} and denoting the right-hand side as $\Lambda'$, we have
\begin{equation}
\begin{split}
    \Lambda'=2\alpha_l\sum_{n_t,n_r}&\mathcal{R}\left\{\text{e}^{j\phi_{l,n_t,n_r}}\sum_is'(\bm{\theta}'_l)[\hat{\bm{x}}_{l}(\hat{\bm{\Theta}}^{\mu})]_{n_t,n_r,i}^*\right\}\\
    &-\alpha_l^2\sum_{n_t,n_r,i}\left|s'(\bm{\theta}'_l)\right|^2,
    \label{eq:likeli}
\end{split}
\end{equation}
where the term $s'(\bm{\theta}'_l)=c^\text{Tx}_{n_t}(\bm{\Omega}^\text{Tx}_{l,n_t})c^\text{Rx}_{n_r}(\bm{\Omega}^\text{Rx}_{l,n_r})R_{\tau_{l,n_t,n_r}}[i]$ is related to the parameter subset $\bm{\theta}'_l=[\tau_l,\varphi^\text{Tx}_l,\theta^\text{Tx}_l,\varphi^\text{Rx}_l,\theta^\text{Rx}_l,d^\text{Tx}_l,d^\text{Rx}_l]$.
\par It can be observed that for any values of $\bm{\theta}'_l$, to maximize the likelihood function, the phase and path gain estimations are given by
\begin{align}
\label{eq:phsest}
    \hat{\phi}_{l,n_t,n_r}&=\text{Angle}\left(\sum_{i=1}s'(\bm{\theta}'_l)[\hat{\bm{x}}_{l}(\hat{\bm{\Theta}}^{\mu})]_{n_t,n_r,i}^*\right),
    \\\hat{\alpha_l}&=\frac{\sum_{n_t,n_r}\left|\sum_{i}s'(\bm{\theta}'_l)[\hat{\bm{x}}_{l}(\hat{\bm{\Theta}}^{\mu})]_{n_t,n_r,i}^*\right|}{\sum_{n_t,n_r}\sum_{i}\left|s'(\bm{\theta}'_l)\right|^2}.
    \label{eq:ampest}
\end{align}
\par By substituting~\eqref{eq:phsest} and~\eqref{eq:ampest} into~\eqref{eq:likeli}, the likelihood function is simplified to
\begin{equation}
    \Lambda'=\frac{\left(\sum_{n_t,n_r}\left|\sum_{i}s'(\bm{\theta}'_l)[\hat{\bm{x}}_{l}(\hat{\bm{\Theta}}^{\mu})]_{n_t,n_r,i}^*\right|\right)^2}{\sum_{n_t,n_r,i}\left|s'(\bm{\theta}'_l)\right|^2}.
\end{equation}
\par Hence, the MLE of the remaining parameters is given as
\begin{align}
    \hat{\bm{\theta}}'_l&=\underset{\bm{\theta}'_l}{{\arg\max}}~\Lambda'.
\end{align}
\par Numerical methods can be applied to find the MLE. However, since scatter distance and DoD/DoA jointly shape the spherical wavefront, the joint estimation of scatter distance and DoD/DoA is required, where an exhaustive search could consume a large amount of time. To overcome this, a coarse-to-fine maximization process is applied as follows.
\subsubsection{Coarse Estimation with PDAP}
\par A key property of the DSS is that the measured PDAP can roughly locate the delay, DoD, and DoA of MPCs. Specifically, as shown in Appendix~\ref{appen:cir}, the maximum of the absolute value of function $R_{\tau_{l}}[i]$ occurs at sample points with delay very close to $\tau_{l}$. Moreover, due to the narrow antenna beam, the Tx/Rx antenna gains are maximum for scanning directions that are closest to $\bm{\Omega}^\text{Tx}_{l}$ and $\bm{\Omega}^\text{Rx}_{l}$. Therefore, assuming that the noise term $\bm{w}'$ is much smaller than the signal $\bm{s}(\bm{\theta}_l)$, coarse estimations can be made as
\begin{align}
    [n_{t,m},n_{r,m},i_m]&=\underset{n_t,n_r,i}{{\arg\max}}([|\hat{\bm{x}}_l(\hat{\bm{\Theta}}^{\mu})|]_{n_t,n_r,i}),
    \label{eq:roughestimation}\\
    \hat{\tau}'_{l}&=(i_m-1)\Delta\tau,\\
    (\hat{\varphi}^\text{Tx}_{l})'&=\varphi^{\text{Tx,Scan}}_{n_{t,m}},\\
    (\hat{\theta}^\text{Tx}_{l})'&=\theta^{\text{Tx,Scan}}_{n_{t,m}},\\
    (\hat{\varphi}^\text{Rx}_{l})'&=\varphi^{\text{Rx,Scan}}_{n_{r,m}},\\
    (\hat{\theta}^\text{Rx}_{l})'&=\theta^{\text{Rx,Scan}}_{n_{r,m}},
    \label{eq:roughestimationdoa}
    \vspace{-0.5cm}
\end{align}
where $\Delta\tau$ denotes the sampling interval in the delay domain. Moreover, $\varphi^{\text{Tx,Scan}}_{n_{t,m}}$ and $\theta^{\text{Tx,Scan}}_{n_{t,m}}$ are the azimuth and elevation angles in the $(n_{t,m})^\text{th}$ scanning direction of Tx, respectively, while $\varphi^{\text{Rx,Scan}}_{n_{r,m}}$ and $\theta^{\text{Rx,Scan}}_{n_{r,m}}$ are the azimuth and elevation angles in the $(n_{r,m})^\text{th}$ scanning directions of Rx, respectively. 
\subsubsection{Fine Search with Partial Data}
\par After obtaining the coarse estimations, we then search for the accurate estimations of parameters of the $l^\text{th}$ MPC in the neighborhood of the coarse estimates. Due to the narrow antenna beam and the form of $R_{\tau_{l}}[i]$, the received signal of the $l^\text{th}$ path is very weak for those scanning directions or temporal samples far from its DoD, DoA, and delay, and may even be buried in noise. Thus, these data do not provide any useful information and can be trimmed to reduce the size of the CIR tensor. By using only the partial data around the coarse estimations, the computational complexity can be greatly reduced. 
\par We proceed by finding the observed delay in the $(n_{t,m})^\text{th}$ Tx scanning direction and $(n_{r,m})^\text{th}$ Rx scanning direction, i.e., the estimation of $\tau_{l,n_{t,m},n_{r,m}}$, as
\begin{equation}
    \label{eq:es_delay}
    \hat{\tau}_{l,n_{t,m},n_{r,m}}=\underset{\tau_l\in\bm{S}_{\tau,l}}{{\arg\max}}\left|\sum_{\bm{I}_{\tau,l}}R^*_{\tau_l}[i][\hat{\bm{x}}_{l}(\hat{\bm{\Theta}}^{\mu})]_{n_{t,m},n_{r,m},i}\right|,
\end{equation}
where $\bm{S}_{\tau,l}$ and $\bm{I}_{\tau,l}$ are the local region and partial data around the coarse delay estimation, as
\begin{align}
    \bm{S}_{\tau,l}&=\left\{\tau;\left|\tau-\hat{\tau}'_{l}\right|<\Delta\tau\right\},\\
    \bm{I}_{\tau,l}&=\left\{i;\left|i-i_m\right|<\Delta I^\text{PI}\right\},
\end{align}
where $\Delta I^\text{PI}$ are the range of the partial data in the delay domain, set according to the measured data.
\par Based on the relationships shown in~\eqref{eq:losdelay} to~\eqref{eq:losdoa}, with the estimation of $\tau_{l,n_{t,m},n_{r,m}}$, the likelihood function is dependent on the DoD, DoA, distance from Tx to FBS, and distance from the Rx to LBS. Therefore, the angle and distance estimations can be found as
\begin{align}
    \label{eq:es_dod}
    [\hat{\varphi}^{\text{Tx},\mu}_{l},\hat{\theta}^{\text{Tx},\mu}_{l},\hat{d}^{\text{Tx},\mu}_l]=~~&\nonumber\\
    \underset{\varphi^\text{Tx}_l\in\bm{S}^\text{Tx}_{\varphi,l},\theta^\text{Tx}_l\in\bm{S}^\text{Tx}_{\theta,l},d^\text{Tx}_l}{{\arg\max}}~&\Lambda'(\bm{\theta}_l;[\hat{\bm{x}}_{l}(\hat{\bm{\Theta}}^{\mu})]_{\bm{N}_{t,l},n_{r,m},\bm{I}_{\tau,l}}),\\
    \label{eq:es_doa}
    [\hat{\varphi}^{\text{Rx},\mu}_{l},\hat{\theta}^{\text{Rx},\mu}_{l},\hat{d}^{\text{Rx},\mu}_l]=~~&\nonumber\\
    \underset{\varphi^\text{Rx}_l\in\bm{S}^\text{Rx}_{\varphi,l},\theta^\text{Rx}_l\in\bm{S}^\text{Rx}_{\theta,l},d^\text{Rx}_l}{{\arg\max}}~&\Lambda'(\bm{\theta}_l;[\hat{\bm{x}}_{l}(\hat{\bm{\Theta}}^{\mu})]_{n_{t,m},\bm{N}_{r,l},\bm{I}_{\tau,l}}),
\end{align}
where $\bm{S}^\text{Tx}_{\varphi,l},\bm{S}^\text{Tx}_{\theta,l},\bm{S}^\text{Rx}_{\varphi,l},\bm{S}^\text{Rx}_{\theta,l}$ in~\eqref{eq:es_dod}-\eqref{eq:es_doa} are the local regions around the coarse estimations, which are expressed as 
\begin{align}
    \bm{S}^\text{Tx}_{\varphi,l}&=\left\{\varphi;\left|\varphi-(\hat{\varphi}^\text{Tx}_{l})'\right|\leq\Delta\varphi^\text{Tx}/2\right\},\\
    \bm{S}^\text{Tx}_{\theta,l}&=\left\{\theta;\left|\theta-(\hat{\theta}^\text{Tx}_{l})'\right|\leq\Delta\theta^\text{Tx}/2\right\},\\
    \bm{S}^\text{Rx}_{\varphi,l}&=\left\{\varphi;\left|\varphi-(\hat{\varphi}^\text{Rx}_{l})'\right|\leq\Delta\varphi^\text{Rx}/2\right\},\\
    \bm{S}^\text{Rx}_{\theta,l}&=\left\{\theta;\left|\theta-(\hat{\theta}^\text{Rx}_{l})'\right|\leq\Delta\theta^\text{Rx}/2\right\},
\end{align}
where $\Delta\varphi^\text{Tx}$ and $\Delta\theta^\text{Tx}$ denotes the angle step for DSS in the azimuth plane and elevation plane for Tx, respectively. Moreover, $\Delta\varphi^\text{Rx}$ and $\Delta\theta^\text{Rx}$ are the angle step for DSS in the azimuth plane and elevation plane for Rx, respectively. Furthermore, $\bm{N}_{t,l}$ and $\bm{N}_{r,l}$ are partial data in the angle domain, expressed as
\begin{align}
    \bm{N}_{t,l}=\{n_t;&\left|\varphi^\text{Tx,Scan}_{n_t}-\varphi^\text{Tx,Scan}_{n_{t,m}}\right|<\Delta\varphi^\text{Tx,PI},\nonumber\\
    &\left|\theta^\text{Tx,Scan}_{n_t}-\theta^\text{Tx,Scan}_{n_{t,m}}\right|<\Delta\theta^\text{Tx,PI}\},\\
    \bm{N}_{r,l}=\{n_r;&\left|\varphi^\text{Rx,Scan}_{n_r}-\varphi^\text{Rx,Scan}_{n_{r,m}}\right|<\Delta\varphi^\text{Rx,PI},\nonumber\\
    &\left|\theta^\text{Rx,Scan}_{n_r}-\theta^\text{Rx,Scan}_{n_{r,m}}\right|<\Delta\theta^\text{Rx,PI}\},
\end{align}
where the thresholds $\Delta\varphi^\text{Tx,PI}, \Delta\theta^\text{Tx,PI}, \Delta\varphi^\text{Rx,PI}, \Delta\theta^\text{Rx,PI}$ are dependent on the beam widths of Tx/Rx antennas, which should be set appropriately to include most of the main beams. 
\par Note that partial data are used to decouple the estimation of DoD and DoD. Specifically, for the estimation of DoD, only the data in the Rx scanning direction receiving the strongest power is selected. Similarly, for the estimation of DoA, only the data in the Tx scanning direction producing the strongest received power is selected. In other words, the estimation problem in the virtual MIMO case is separated into two problems in the MISO and SIMO cases.
\par After obtaining the estimations of DoD, DoA and scatter distances of the $l^\text{th}$ MPC, the estimation of delay, path gain, and phase can be obtained as
\begin{align}
    \label{eq:es_tau}
    \hat{\tau}^\mu_l&=\hat{\tau}_{l,n_{t,m},n_{r,m}}-\frac{||\hat{\bm{r}}^{\text{FBS}}-\bm{r}^{\text{Tx}}_{n_{t,m}}||-\hat{d}^{\text{Tx},\mu}_{l}}{c}\nonumber\\
    &~~~~~~~~~~~~~~~~~-\frac{||\hat{\bm{r}}^{\text{LBS}}-\bm{r}^{\text{Rx}}_{n_{r,m}}||-\hat{d}^{\text{Rx},\mu}_{l}}{c},\\
    \hat{\alpha}^\mu_l&=\frac{\sum\limits_{\bm{N}_{t,l}, \bm{N}_{r,l}}c^\text{Tx}_{n_{t}}(\hat{\bm{\Omega}}^{\text{Tx},\mu}_{l})c^\text{Rx}_{n_{r}}(\hat{\bm{\Omega}}^{\text{Rx},\mu}_{l})\left|\hat{\alpha}_{l,n_t,n_r}\right|}{\sum\limits_{\bm{N}_{t,l}, \bm{N}_{r,l},\bm{I}_{\tau,l}}c^\text{Tx}_{n_{t}}(\hat{\bm{\Omega}}^{\text{Tx},\mu}_{l})c^\text{Rx}_{n_{r}}(\hat{\bm{\Omega}}^{\text{Rx},\mu}_{l})\left|R_{\tau_{l,n_t,n_r}}[i])\right|^2},\\
    &\hat{\phi}^\mu_{l,n_t,n_r}=\text{Angle}(\hat{\alpha}_{l,n_t,n_r}),
\end{align}
where the positions of FBS and LBS are estimated as
\begin{align}
    \hat{\bm{r}}^{\text{FBS}}=\overline{\bm{r}}^\text{Tx}+\hat{d}^{\text{Tx},\mu}_l\hat{\bm{\Omega}}^{\text{Tx},\mu}_l,\\
    \hat{\bm{r}}^{\text{LBS}}=\overline{\bm{r}}^\text{Rx}+\hat{d}^{\text{Rx},\mu}_l\hat{\bm{\Omega}}^{\text{Rx},\mu}_l,
\end{align}
Moreover, the observed signal in the $(n_t)^\text{th}$ scanning direction of Tx and $(n_r)^\text{th}$ scanning direction of Rx is given as
\begin{align}
    \hat{\alpha}_{l,n_t,n_r}&=\sum\limits_{i\in\bm{I}_{\tau,l}}\left[\hat{\bm{x}}_{l}(\hat{\bm{\Theta}}^{\mu})\right]_{n_t,n_{r},i}R^*_{\tau_{l,n_t, n_r}}[i])
\end{align}
\par Upon here, one iteration of the DSS-o-SAGE algorithm is finished and the parameter of the $l^\text{th}$ MPC is updated. 
\subsection{Iteration Cycle and Convergence Judgement}
\label{sec:outerloop}
\par Each $L$ iteration of the DSS-o-SAGE algorithm forms an iteration cycle. The number of MPCs is determined in the first iteration cycle, namely the initialization cycle, which starts with all MPC parameters set to zero. During the initialization cycle, the MPCs are estimated and subtracted one by one, in the descending order in light of path gains, until the estimated path gain of the $l^\text{th}$ path is smaller than a certain threshold. The number of MPCs is then set as the counter $l$ when the initialization cycle terminates.
\par At the end of each iteration cycle, i.e., $\mu \text{mod} (L)=L-1$, we judge whether the algorithm converges by comparing the values of the likelihood function in~\eqref{eq:likeliall}. Specifically, if the increase in the likelihood function compared to that in the last iteration cycle is less than a threshold, the algorithm is considered to be converged, expressed as
\begin{equation}
    \Lambda(\hat{\bm{\Theta}}^\mu;\bm{h})-\Lambda(\hat{\bm{\Theta}}^{\mu-L};\bm{h})<r\Lambda(\hat{\bm{\Theta}}^{\mu-L};\bm{h})
    \label{eq:converge}
\end{equation}
where $r$ denotes the ratio threshold.
\par What's more, the algorithm can also be terminated when the number of iteration cycles exceeds a certain number, e.g. 10 in this work.
\section{Performance Validation and Comparison in Synthetic Channels}
\label{sec:com}
\par In this section, the performance of the proposed DSS-o-SAGE algorithm is validated and compared with existing methods considering synthetic channels, including the SAGE algorithm based on PWF (called PWF-SAGE in this work to distinguish it from others) and the SWF-SAGE algorithm~\cite{fleury1999channel,yin2017scatter}. The computational complexity of different estimation algorithms is compared by evaluating the running time of a M-step. Furthermore, under phase instability in mmWave and THz DSS, results using different algorithms as well as the Cram\'er-Rao lower bound are compared and discussed. {Last but not least, the influence of the near-field on estimation performance of SAGE algorithms is studied, where the applicable region of far-field approximations in DSS-o-SAGE is modeled.}
\subsection{{Synthetic Channel - Single Path Case}}
\label{sec:simulation}
\par To evaluate the performance of the proposed estimation algorithm, simulations are conducted with synthetic channels. Without loss of generality, we consider a single-input-multiple-output (SIMO) case where only the Rx scans the spatial domain and only the DoA is estimated while the DoD is omitted. The scanning direction of Rx ranges from $0^\circ$ to $350^\circ$ in the azimuth plane and from $-20^\circ$ to $20^\circ$ in the elevation plane, both with a step of $10^\circ$. Moreover, it is assumed that the CIR data is measured with the vector network analyzer (VNA)-based channel sounder. The frequency band is selected as \SIrange{298}{302}{GHz}, with a \SI{2}{MHz} sampling interval. Furthermore, the horizontal and azimuth radii of the rotator is assumed to be $\sqrt{0.02}$, for which the antenna aperture is $2R=$\SI{0.4}{m} and the corresponding Rayleigh distance is \SI{320}{m}.
\begin{table*}[tbp]
    \centering
    \caption{Computational complexity and wall time in one M-step of different SAGE algorithms.}
    \includegraphics[width = 1.9\columnwidth]{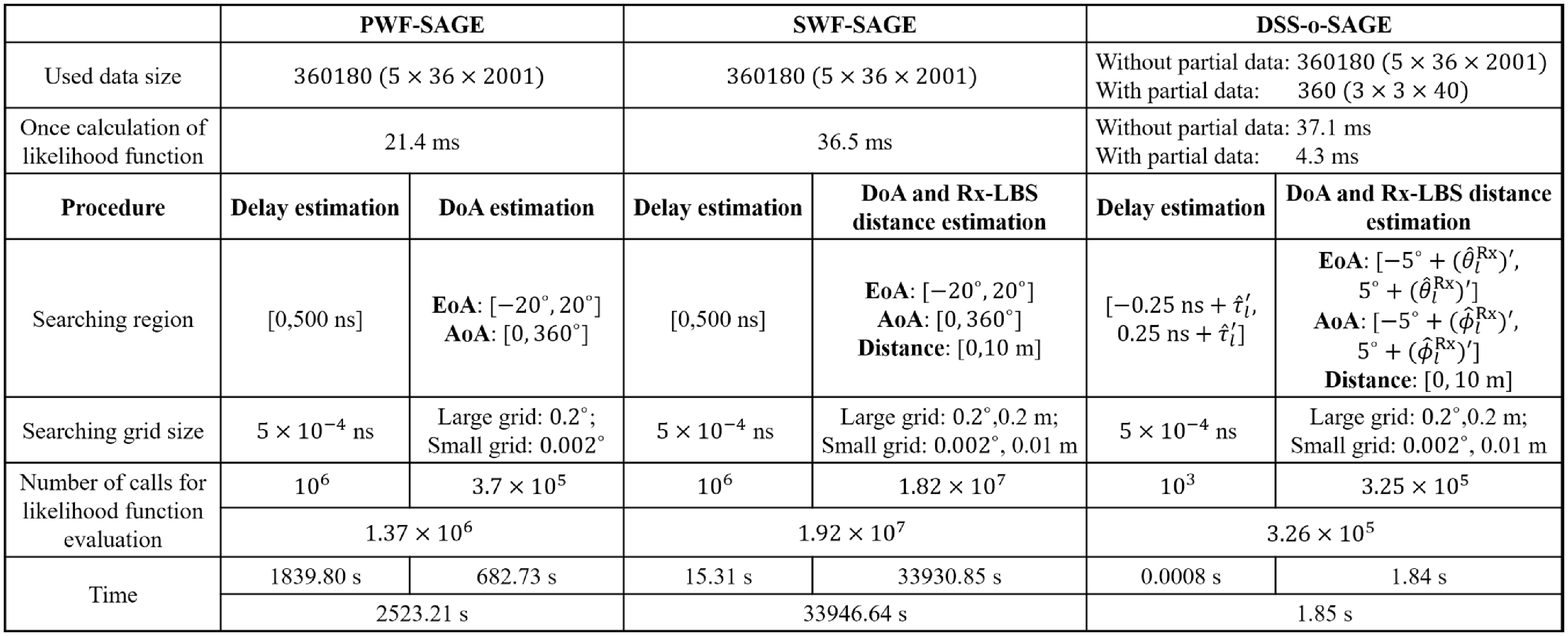}
    \label{tab:complex}
    \vspace{-0.3 cm}
\end{table*}
\par {To compare the performance of different algorithms, a single path case is considered, where there is only one LoS path in the channels. The parameter of the path is given as}
\begin{align}
    [\tau_1,\varphi^\text{Rx}_1,\theta^\text{Rx}_1,d^\text{Rx}_1,\alpha_1]&=[\frac{d}{c},\frac{\Delta\varphi}{2},\frac{\Delta\theta}{2},d,\frac{c}{4\pi fd}],
\end{align}
{where $d$ denotes the propagation distance. Moreover, $\Delta\varphi$ and $\Delta\theta$ are scanning steps in the azimuth and elevation plane. In other words, the LoS path is assumed to be arrived in directions in the middle of scanning grids, which is the hardest case for parameter estimation. Furthermore, the path gain is set according to the Friss' law. Apart from the parameters listed above, the signal phases in different scanning directions are generated following zero-mean Gaussian distributions.}
\par With the aforementioned parameters, the measured CIR tensor can be constructed following~\eqref{eq:cirtensor}, where the noise samples are generated with complex normal distributions.
\subsection{Computational Complexity}
\par All three SAGE algorithms find the estimations through on-grid search process. Specifically, for the PWF-SAGE algorithm, the delay is estimated with a grid size of $5\times10^{-4}$~ns. Moreover, the AoA and EoA are jointly estimated with a two-phase on-grid search. A large grid with grid size of $0.2^\circ$ is firstly used to find the coarse estimations, within whose neighbourhood a small grid with size of $0.002^\circ$ is then used to find the final estimation values. Moreover, for the SWF-SAGE algorithm, the delay is firstly estimated with a grid size of $5\times10^{-4}$~ns, by using only the data in the strongest scanning direction. Then, the AoA, EoA, and distance from Rx to LBS are jointly estimated, firstly with large grid sized as $0.2^\circ$ and \SI{0.2}{m} and then with small grid sized as $0.002^\circ$ and \SI{0.01}{m}. Similarly, the delay in DSS-o-SAGE is estimated with a grid size of $5\times10^{-4}$~ns. The AoA, EoA, and Rx-LBS distance are estimated with large grids separated by $0.2^\circ$ and \SI{0.2}{m} and small grids separated by $0.002^\circ$ and \SI{0.01}{m}.
\par The computational complexity and wall time in one M-step for different SAGE algorithms are shown in Table~\ref{tab:complex}, from which several observations can be made as follows. Note that the wall times are recorded using the same implementation environment and computer specifications. {First, comparing the wall time for once calculation of the likelihood function, the PWF-SAGE algorithm is faster than the SWF-SAGE algorithm, since the calculation of SWF is more complex than PWF. Moreover, without using partial data, the wall time for once calculation of the likelihood function in DSS-o-SAGE algorithm is close to that in the SWF-SAGE algorithm. By utilizing the partial data, the involved data size can be reduced by about 1000 times and nearly an order of magnitude decrease is obtained for the wall time of the likelihood function evaluation, which validates the effective computational acceleration with the proposed partial data utilization.} Second, the number of calls for likelihood function evaluation in the DSS-o-SAGE algorithm is much smaller than that in the other two algorithms, with a decrease of one order compared to PWF-SAGE algorithm and two orders compared to SWF-SAGE algorithm. This attributes to the coarse estimations obtained utilizing the narrow beam property of the DSS. In summary, the combination of the smaller data size and smaller search space results in the much smaller wall time in the DSS-o-SAGE algorithm, compared to the other two algorithms.
\par As shown in Table~\ref{tab:complex}, the computational complexity without using coarse estimation and partial data in PWF-SAGE and SWF-SAGE is too high, which prevents us from running Monte Carlo simulations to evaluate their performance. Thus, for our simulations, these two algorithms are also accelerated using the coarse estimation and partial data.
\subsection{Influences of Phase Instability}
\label{sec:accu}
\par To compare the estimation performance of different SAGE algorithms, the estimation accuracy is evaluated by the root-mean-square error (RMSE), which is expressed as
\begin{align}
    \text{RMSE}(\Theta_k)=(\Theta_k^\text{E}-\Theta_k^\text{T})^2,
\end{align}
where $\Theta_k^\text{E}$ and $\Theta_k^\text{E}$ stands for the estimated parameters and the true values. Moreover, the estimation errors of ToA, DoA, and path gain are considered in this work, i.e., $\Theta_k\in\{\tau,\varphi,\theta,\alpha\}$.
\par {Apart from the estimation accuracy, another important issue is whether the estimation algorithms would produce fake estimations. To investigate this, after the single LoS path is estimated, its effects are subtracted from the channel data and a fake path is then estimated. As a result, the fake power ratio (FPR) is calculated as
\begin{equation}
    r_\text{Fake} [\text{dB}]=20\log_{10}\frac{\alpha_{\text{Fake}}}{\alpha}
\end{equation}
where $\alpha_{\text{Fake}}$ denotes the path gain of the fake estimation.}
\par Moreover, for an unbiased estimator $\hat{\bm{\Theta}}_k$ of the $k^\text{th}$ parameter in $\bm{\Theta}$, its RMSE is lower bounded by the square root of the CRLB, which is defined as
\begin{equation}
    \text{CRLB}(\bm{\Theta}_k)\triangleq[\bm{F}^{-1}(\bm{\Theta})]_{kk}
\end{equation}
where $\bm{F}^{-1}(\bm{\Theta})$ denotes the inverse matrix of the the Fisher Information Matrix (FIM), whose element at the $k^\text{th}$ row and $(k')^\text{th}$ column is expressed as
\begin{equation}
    F_{kk'}(\bm{\Theta})=-\bm{E}\left[\frac{\partial}{\partial\bm{\Theta}_k}\frac{\partial}{\partial\bm{\Theta}_{k'}}\Lambda(\bm{\Theta};\bm{h})\right],
\end{equation}
where $\Lambda(\bm{\Theta};\bm{h})$ is the likelihood function. 
\par With the signal form expressed in~\eqref{eq:cirtensor} and assuming that the noise samples are independent complex Gaussian random variables, the element of FIM can be easily derived to be~\cite{fleury1999channel}
\begin{equation}
    F_{kk'}(\bm{\Theta})=\frac{2}{N_0}\mathcal{R}\left\{\frac{\partial\text{vec}[\bm{s}(\bm{\Theta})]^\text{T}}{\partial\bm{\Theta}_k}\frac{\partial\text{vec}[\bm{s}(\bm{\Theta})]}{\partial\bm{\Theta}_{k'}}\right\}
    \label{eq:fim}
\end{equation}
where $(\cdot)^\text{T}$ denotes the transpose operation.
\par Since there is no analytical expression for the radiation pattern of Tx/Rx horn antennas, the closed-form derivation of the CRLB is not applicable in this work. Nevertheless, the FIM can be easily calculated with numerical solutions based on~\eqref{eq:fim} since it only involves the calculation of first-order derivatives. As a result, the CRLB is computed numerically. 
{Note that only the CRLB with the signal model introduced in this work is evaluated here, as a reference to observe how good or bad the different SAGE algorithms are. Though misspecified CRLBs with the mismatched signal models in PWF-SAGE and SWF-SAGE algorithms are tighter lower bounds~\cite{Richmond2015Parameter}, their evaluations are complicated and beyond the scope of this paper, which are thus omitted.}
\label{sec:spsimulation}
\begin{figure}[!tbp]
\centering
\subfloat[Path gain.]{
\includegraphics[width=0.24\textwidth]{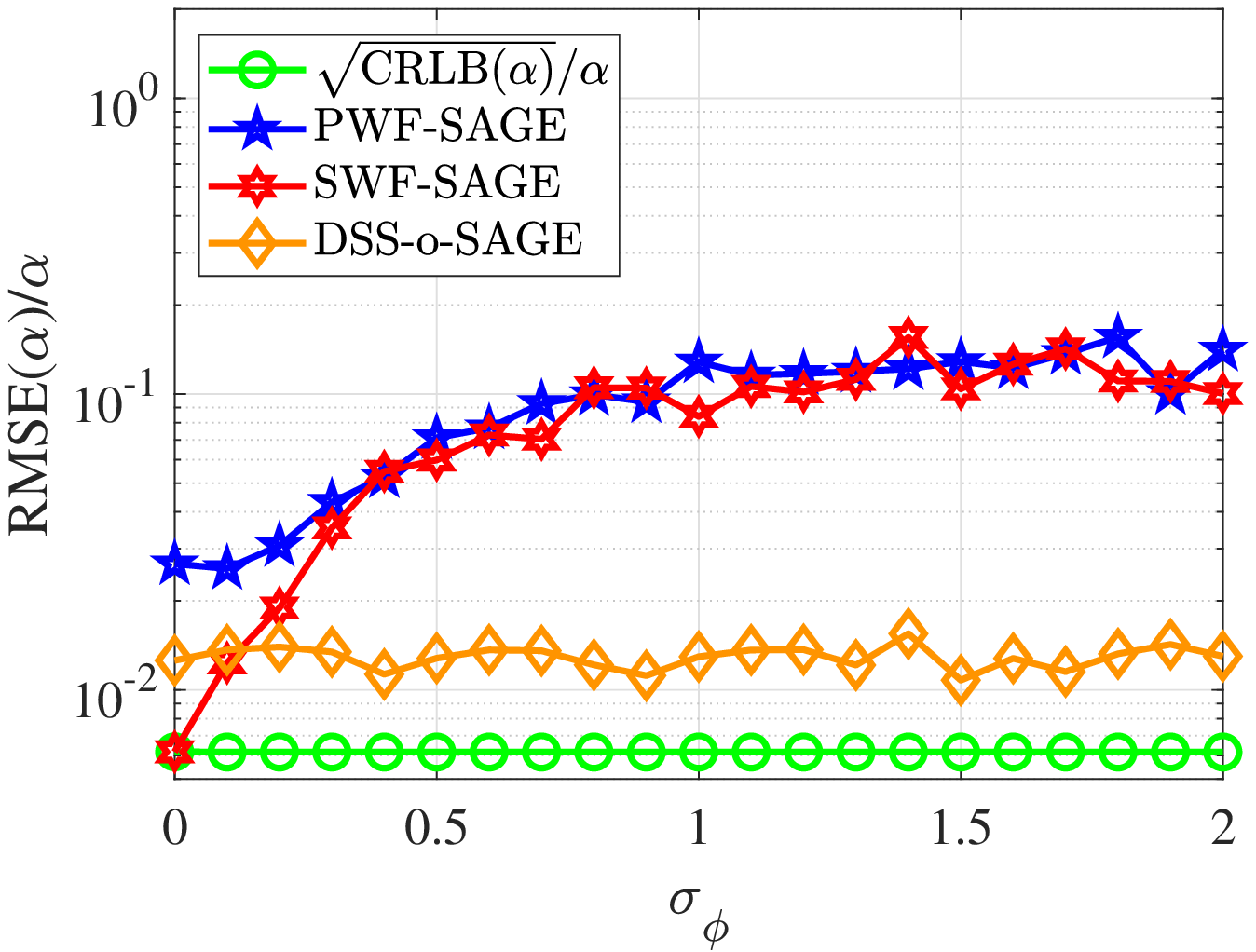}
}
\subfloat[ToA.]{
\includegraphics[width=0.24\textwidth]{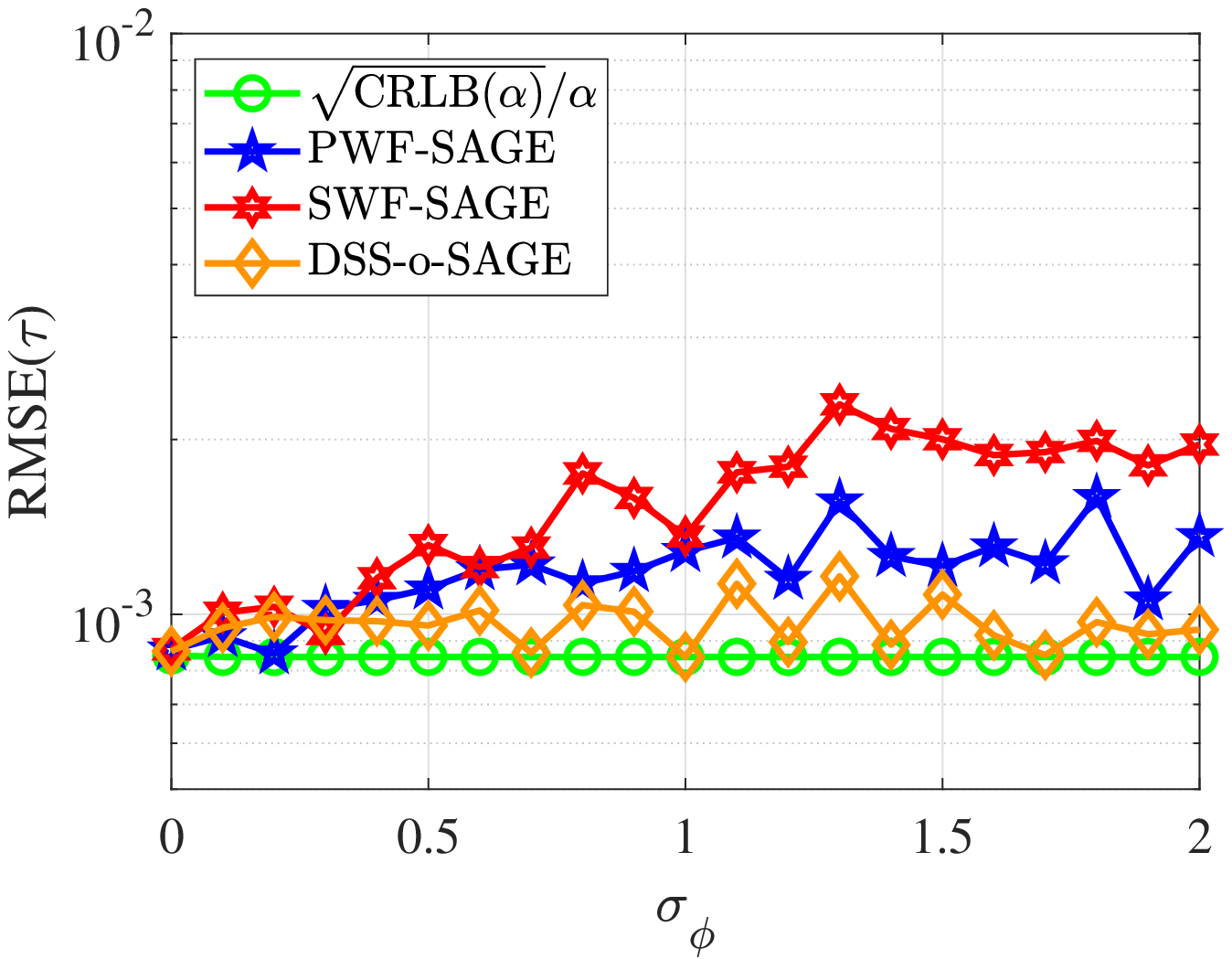}
}
\quad
\subfloat[AoA.]{
\includegraphics[width=0.24\textwidth]{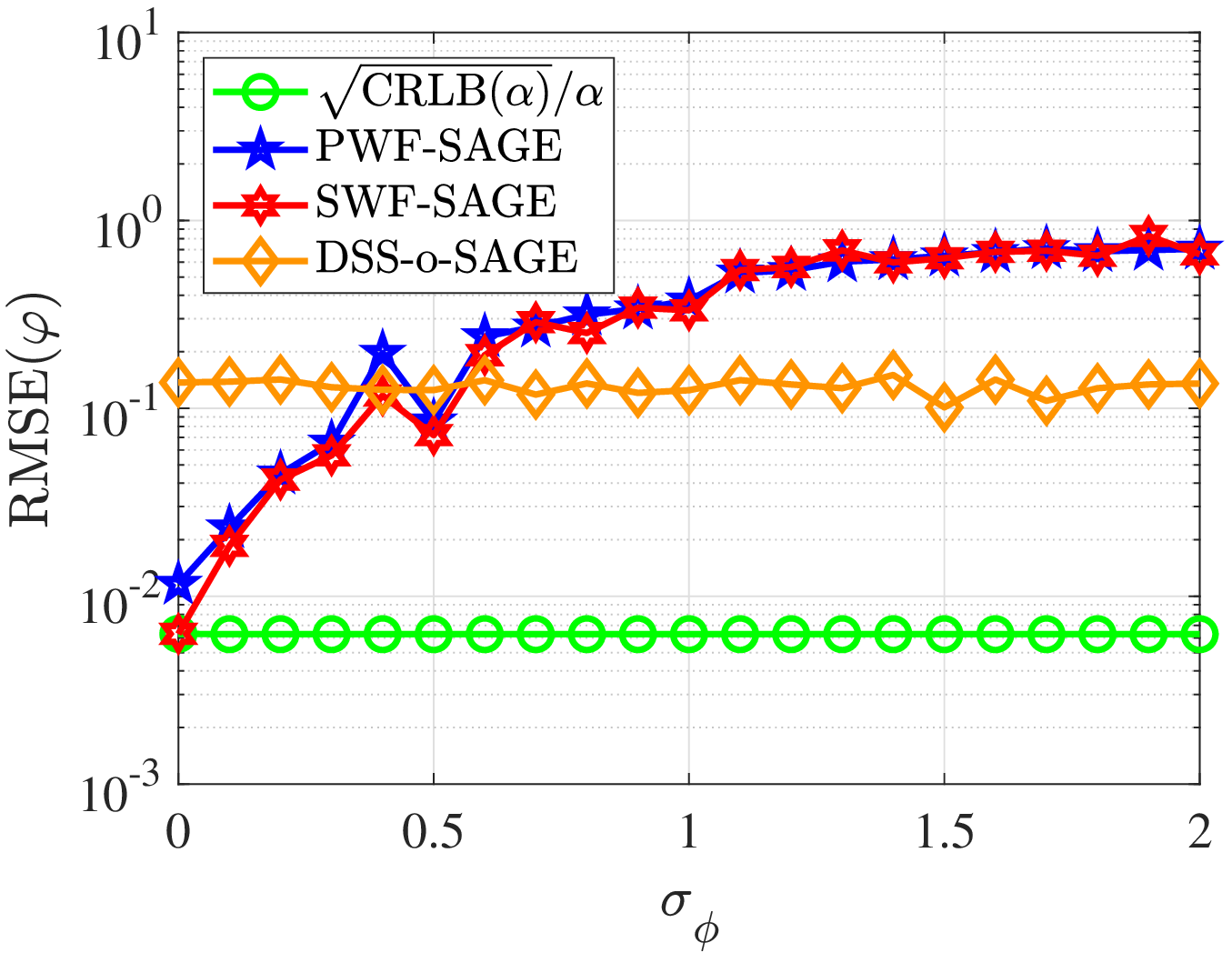}
}
\subfloat[EoA.]{
\includegraphics[width=0.24\textwidth]{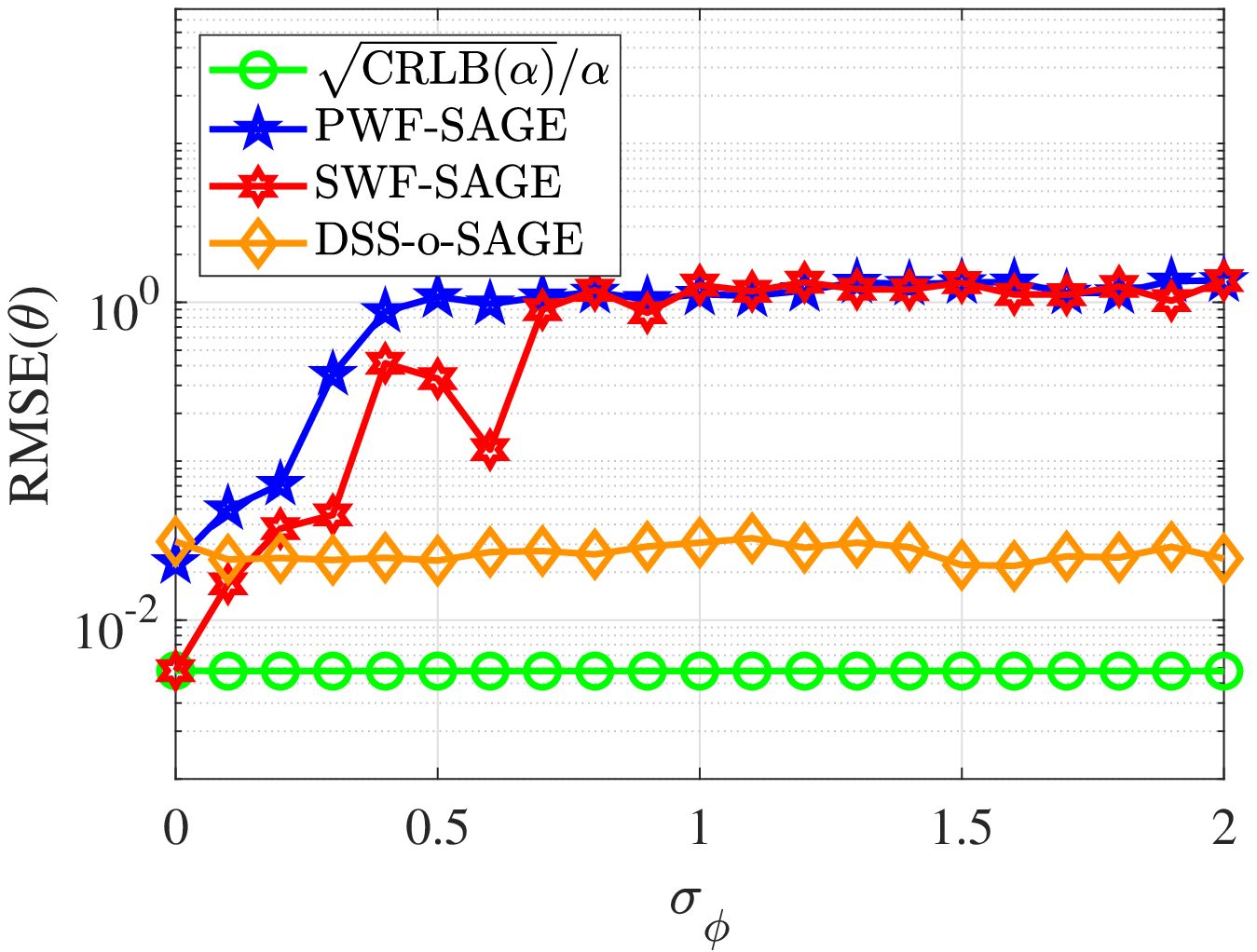}
}
\quad
\subfloat[FPR.]{
\includegraphics[width=0.24\textwidth]{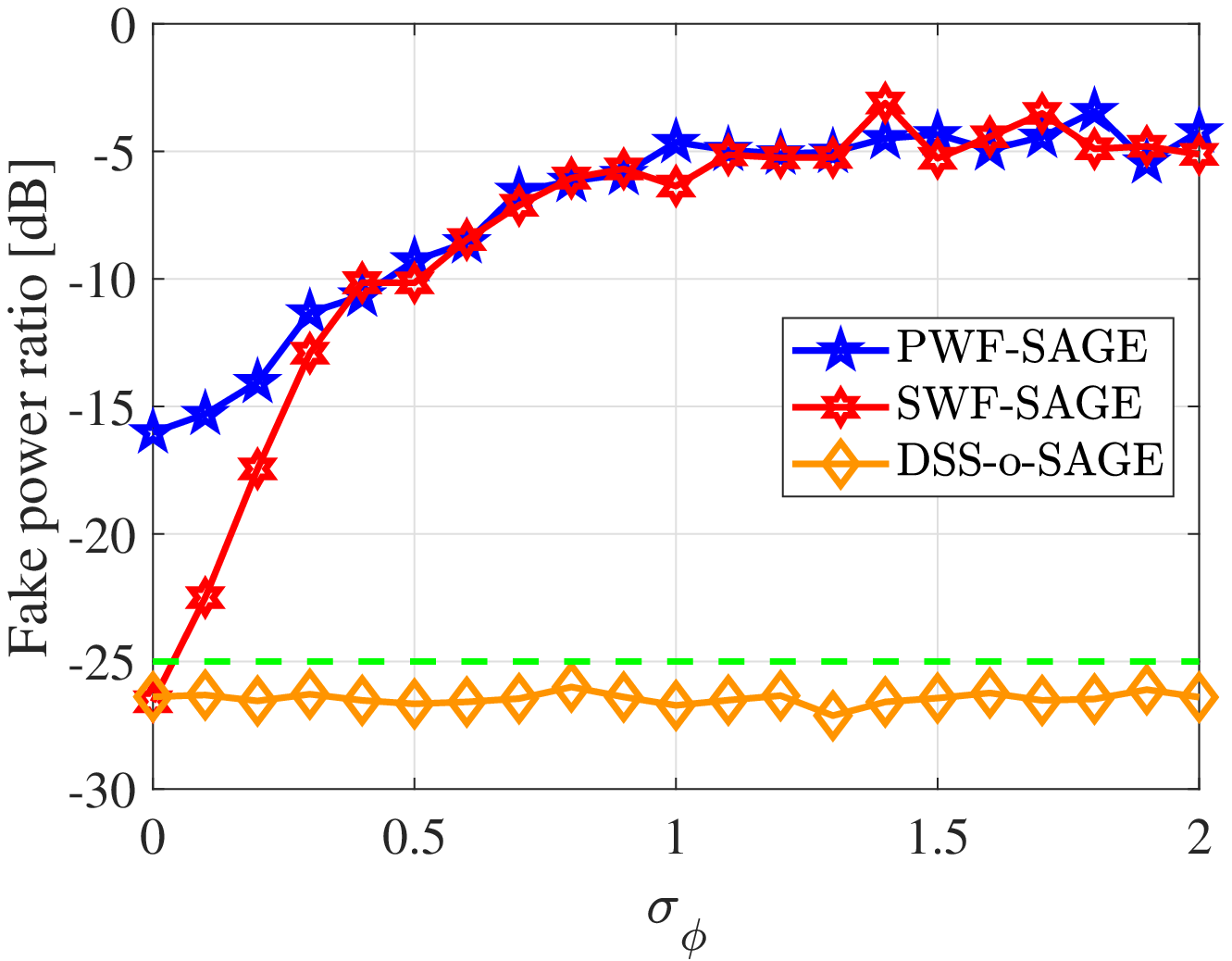}
}
\caption{{Comparison of estimation performance of different SAGE algorithms against phase instability.}}
\label{fig:phasevary}
\vspace{-0.5 cm}
\end{figure}
\par {To study the influence of phase instability, the scanning-direction-dependent phases are generated following zero-mean normal distributions with standard deviations ranging from 0 to 2, representing the cases from no phase instability to severe phase instability as observed in our experiment in Sec.II-B. The distance is set as \SI{10}{m} and the SNR is selected as \SI{40}{dB}. The simulation is run repeatedly for 100 times.}

{The simulation results under phase instabilities are shown in Fig.~\ref{fig:phasevary}. First, when the phase instability is insignificant, e.g., $\sigma_\phi=0$, the estimation accuracy of SWF-SAGE reaches the CRLB. In contrast, the estimation errors using PWF-SAGE algorithm are larger, due to the near-field effects under the considered \SI{10}{m} scatter distance. Second, the PWF-SAGE and SWF-SAGE algorithms suffer greatly from the phase instability. As the phase instability is severer, their estimation accuracy degrades. When phase instability is significant, e.g., $\sigma_\phi>1$, the PWF-SAGE and SWF-SAGE algorithms can not work at all and produce strong fake paths. Therefore, under the severe phase instability as observed in our experiments in Sec. II-B, namely $\sigma_\phi\approx1.8$, the existing algorithms can not be used for accurate channel parameter estimations. Third, with the carefully modeled scanning-direction-dependent signal phases, the performance of the DSS-o-SAGE algorithm is robust under phase instability. These results show the excellence of DSS-o-SAGE compared to existing algorithms.}
\begin{figure}[!tbp]
\centering
\subfloat[Path gain.]{
\includegraphics[width=0.24\textwidth]{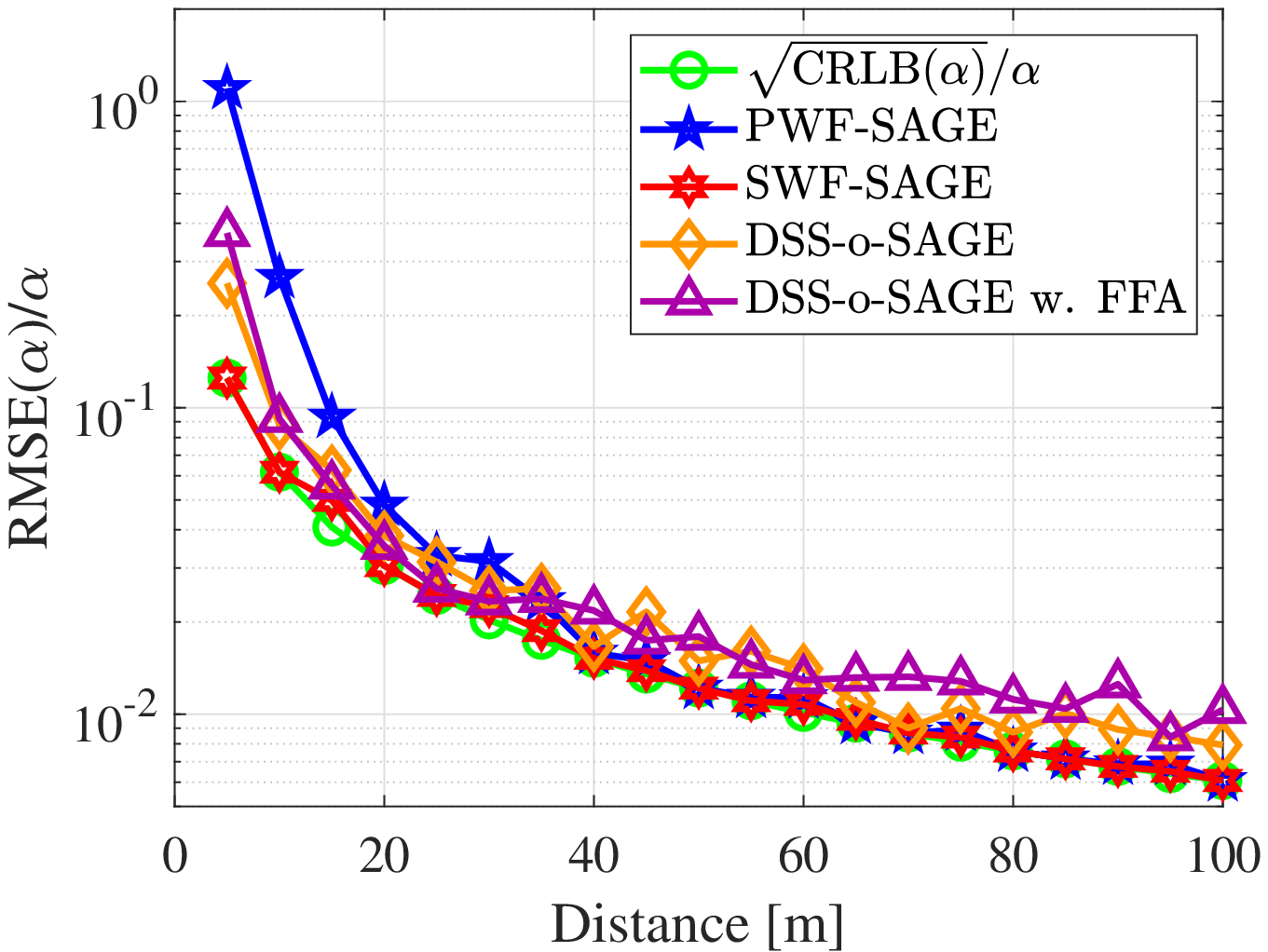}
}
\subfloat[ToA.]{
\includegraphics[width=0.24\textwidth]{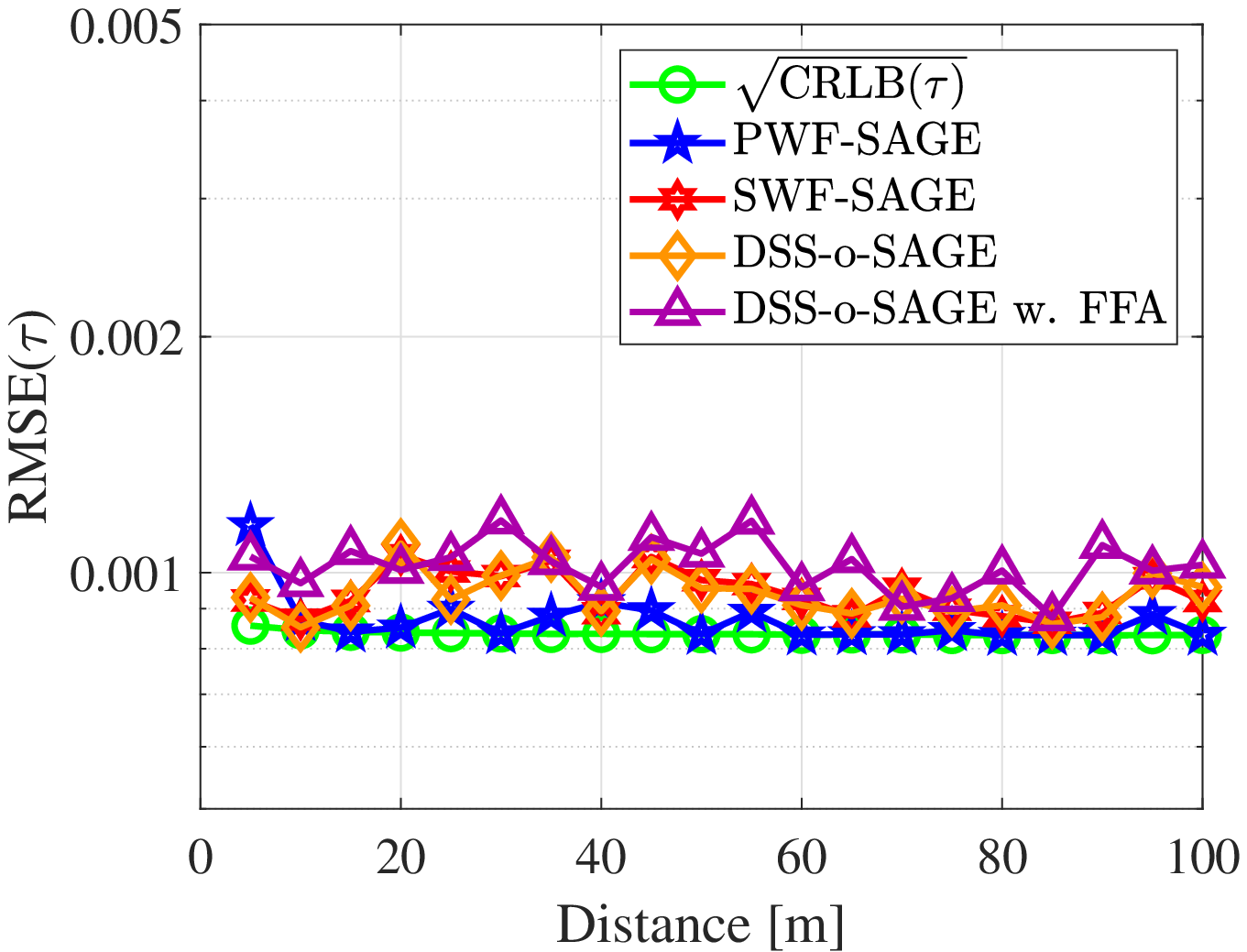}
}
\quad
\subfloat[AoA.]{
\includegraphics[width=0.24\textwidth]{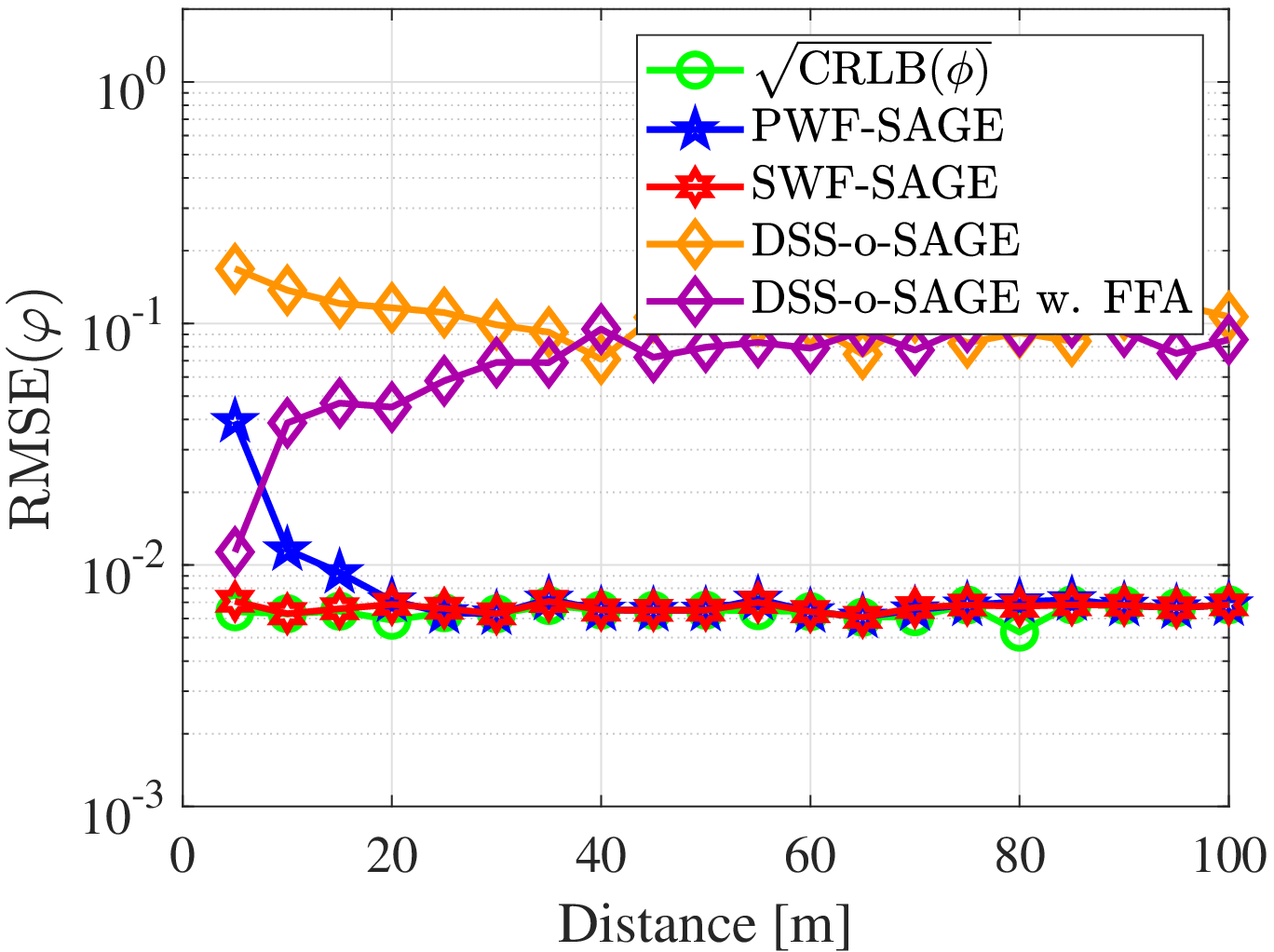}
}
\subfloat[EoA.]{
\includegraphics[width=0.24\textwidth]{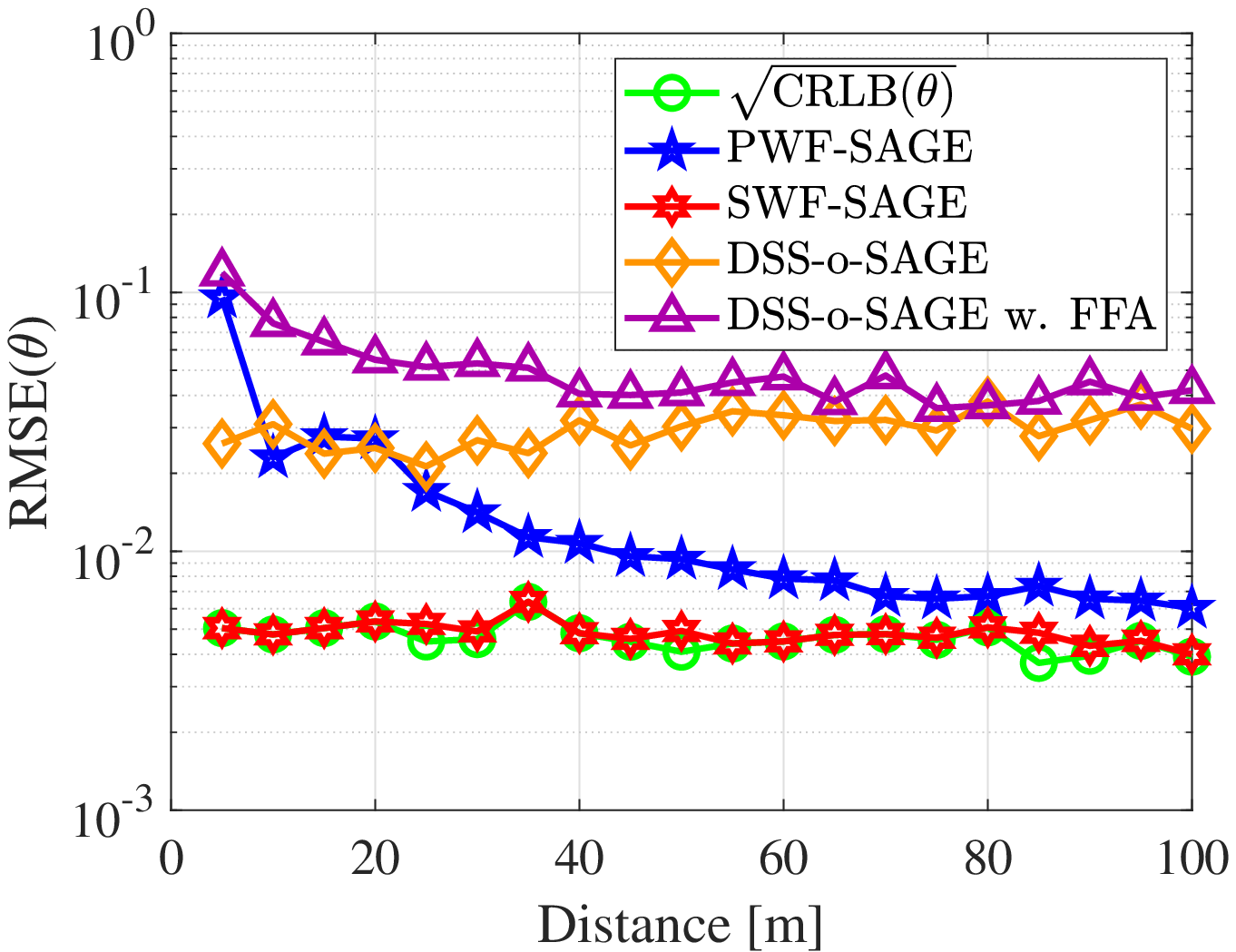}
}
\quad
\subfloat[FPR.]{
\includegraphics[width=0.24\textwidth]{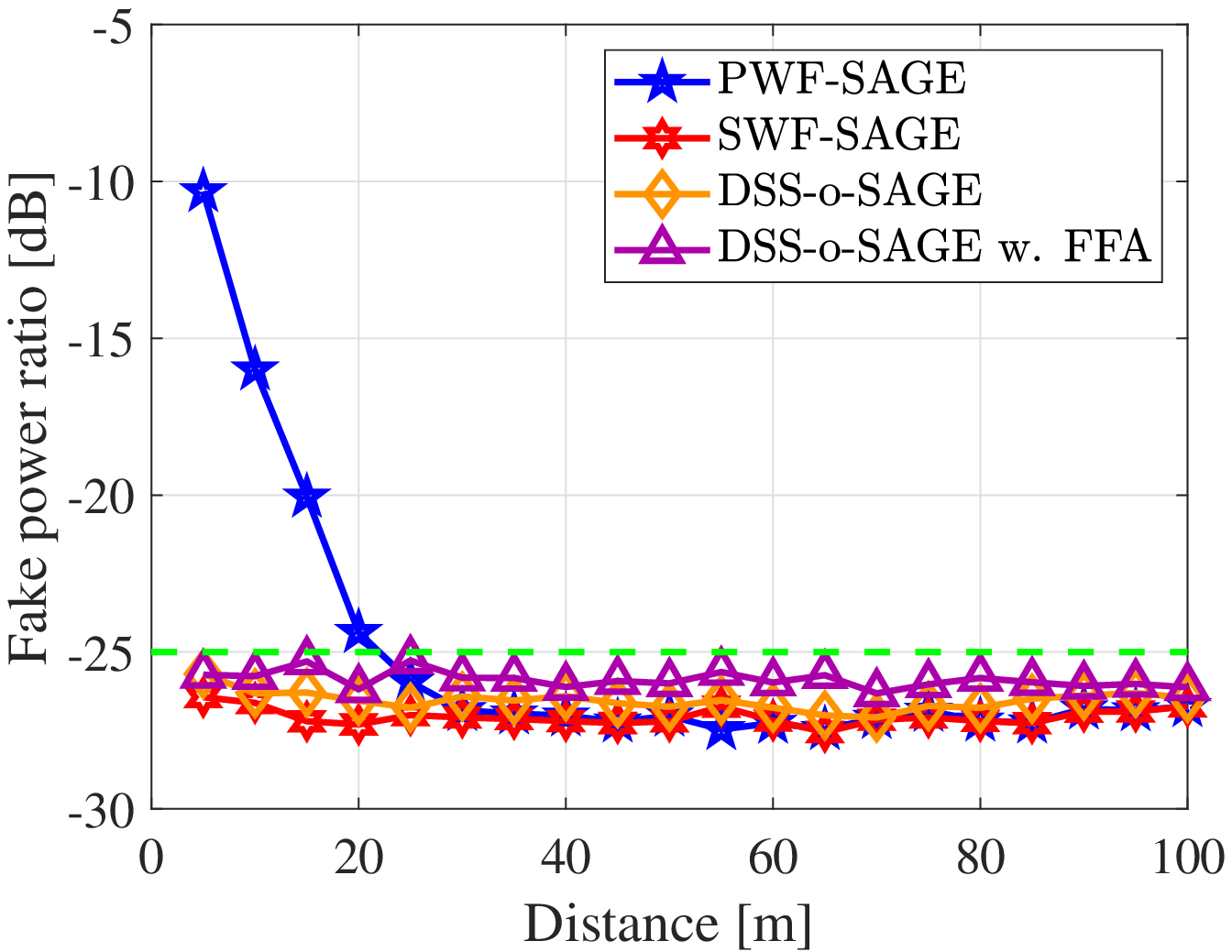}
}
\caption{{Comparison of estimation performance of different SAGE algorithms against scatter distance.}}
\label{fig:disvary}
\vspace{-0.5 cm}
\end{figure}
\subsection{{Influence of Near-field}}
\label{sec:nearfield}
\par {One key concern in mmWave and THz DSS is that whether the large VSA aperture and small wavelength will introduce strong near-field effects that need to be considered for channel parameter estimation. To investigate this, simulations are performed with different scatter distances, where the results are shown in Fig.~\ref{fig:disvary}. No phase instability is considered and the SNR is set as \SI{40}{dB}. Note that to investigate how the performance of DSS-o-SAGE algorithm is affected by the near-field effects, results of DSS-o-SAGE algorithm with far-field approximation (DSS-o-SAGE w. FFA) are also shown here, which can be easily obtained by letting scatter distance of all MPCs approach infinity in the DSS-o-SAGE algorithm.}

{Several observations are made as follows. First, since the phase instability is not considered, the SWF-SAGE algorithm performs well, with estimation accuracy approaching the CRLB. Second, for distances smaller than \SI{20}{m}, the estimation error of the PWF-SAGE algorithm deviates from the CRLB, due to the near-field effects. As the distance increases, the estimation error of the PWF-SAGE algorithm approaches the CRLB. Interestingly, the \SI{20}{m} boundary, above which the estimation accuracy of PWF-SAGE algorithm is satisfactory, is much smaller than the Rayleigh distance of \SI{320}{m}. This is due to the high directivity of the horn antennas used in the DSS, for which the scanning directions with significant model mismatch receive insignificant power, causing little influence on the estimation accuracy. Third, the estimation error of the DSS-o-SAGE algorithm is slightly larger than the CRLB, especially for the angle estimations. Nonetheless, the estimation accuracy is still satisfactory. Fourth, the estimation accuracy of DSS-o-SAGE w. FFA is close to that of the DSS-o-SAGE algorithm, indicating that the DSS-o-SAGE algorithm is less sensitive to the near-field effects than the PWF-SAGE algorithm. The reason for the slightly larger estimation errors and the insensitivity to near-field effects originates from the less information in DSS-o-SAGE algorithm, compared to existing algorithms, as independent signal phases in different scanning directions are considered. Interestingly, the AoA estimation accuracy of the DSS-o-SAGE w. FFA is even smaller than that of the DSS-o-SAGE algorithm when distance is smaller than \SI{40}{m}. Last but not least, as can be observed in Fig.~\ref{fig:disvary}(e), the DSS-o-SAGE algorithm, DSS-o-SAGE w. FFA, and SWF-SAGE algorithms produces weak fake estimations. In fact, the fake estimations produced in these methods are from the noise effects, which can be simply eliminated by selecting a proper threshold, e.g., $r_{\text{Fake},\text{th}}=-25\text{dB}$. In contrast, when distance is smaller than \SI{20}{m}, the PWF-SAGE algorithm produces significant fake MPCs, which prevents its effective use.}
\begin{figure}[!tbp]
\centering
\subfloat[$5^\circ$.]{
\includegraphics[width=0.24\textwidth]{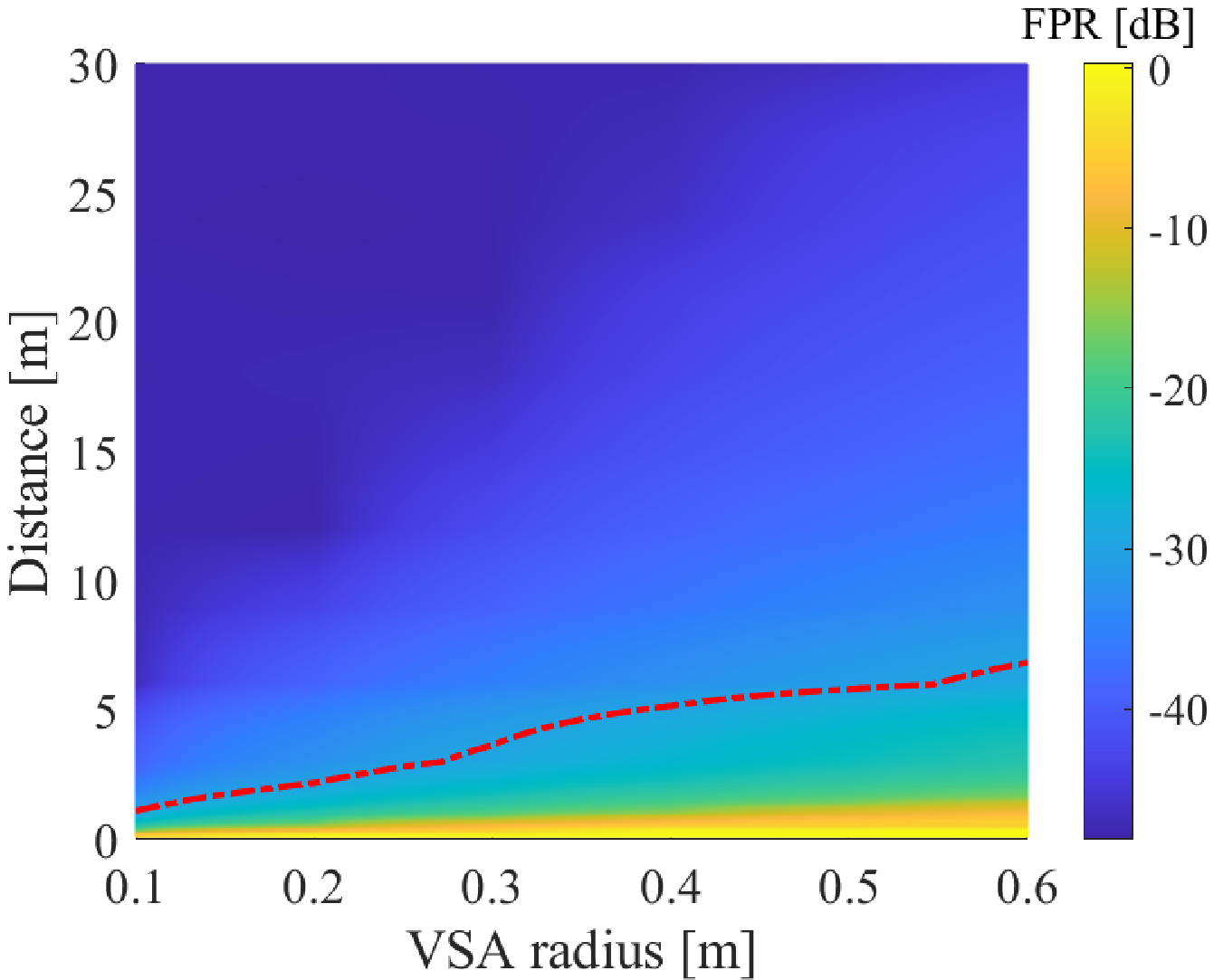}
}
\subfloat[$10^\circ$.]{
\includegraphics[width=0.24\textwidth]{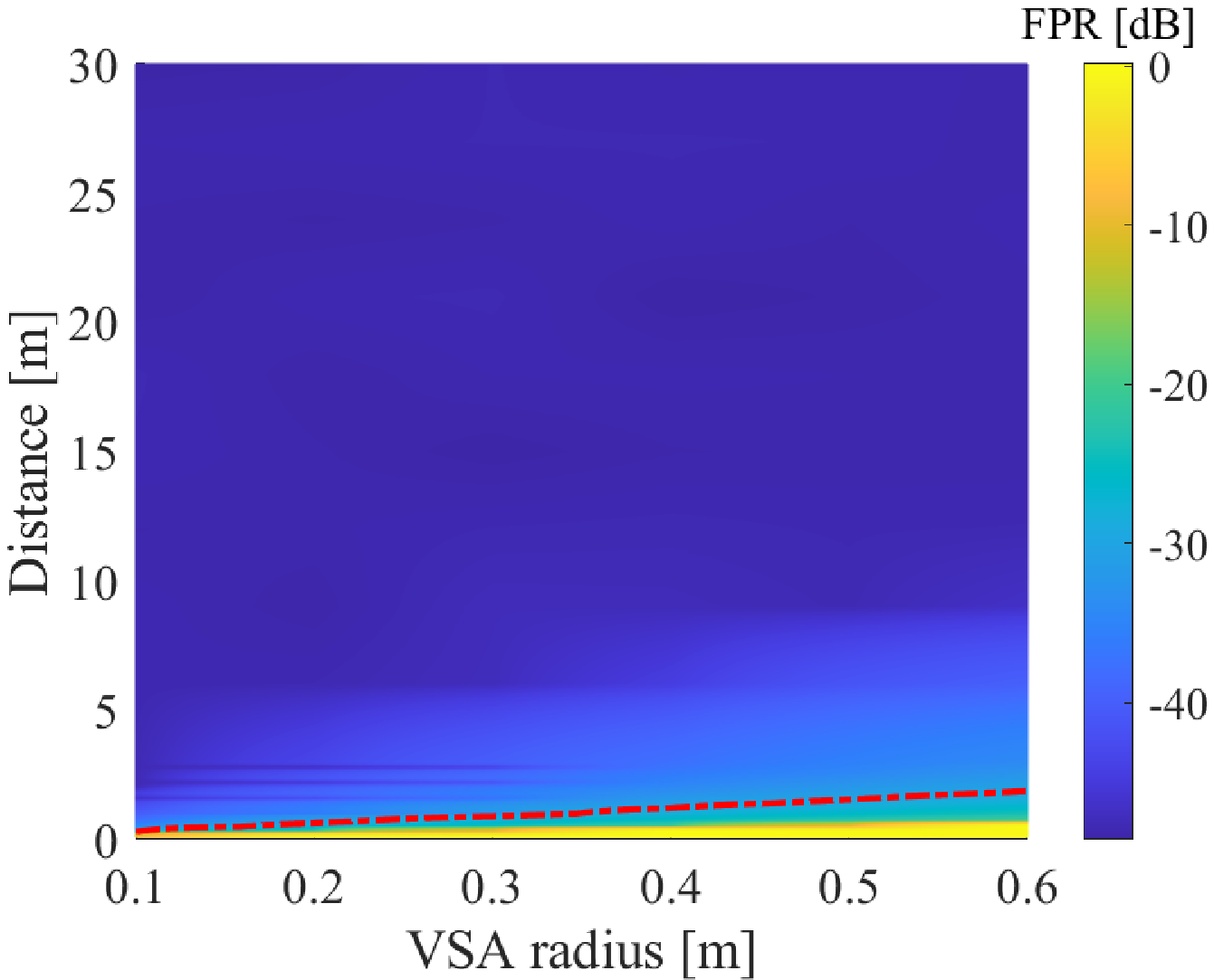}
}
\quad
\subfloat[$20^\circ$.]{
\includegraphics[width=0.24\textwidth]{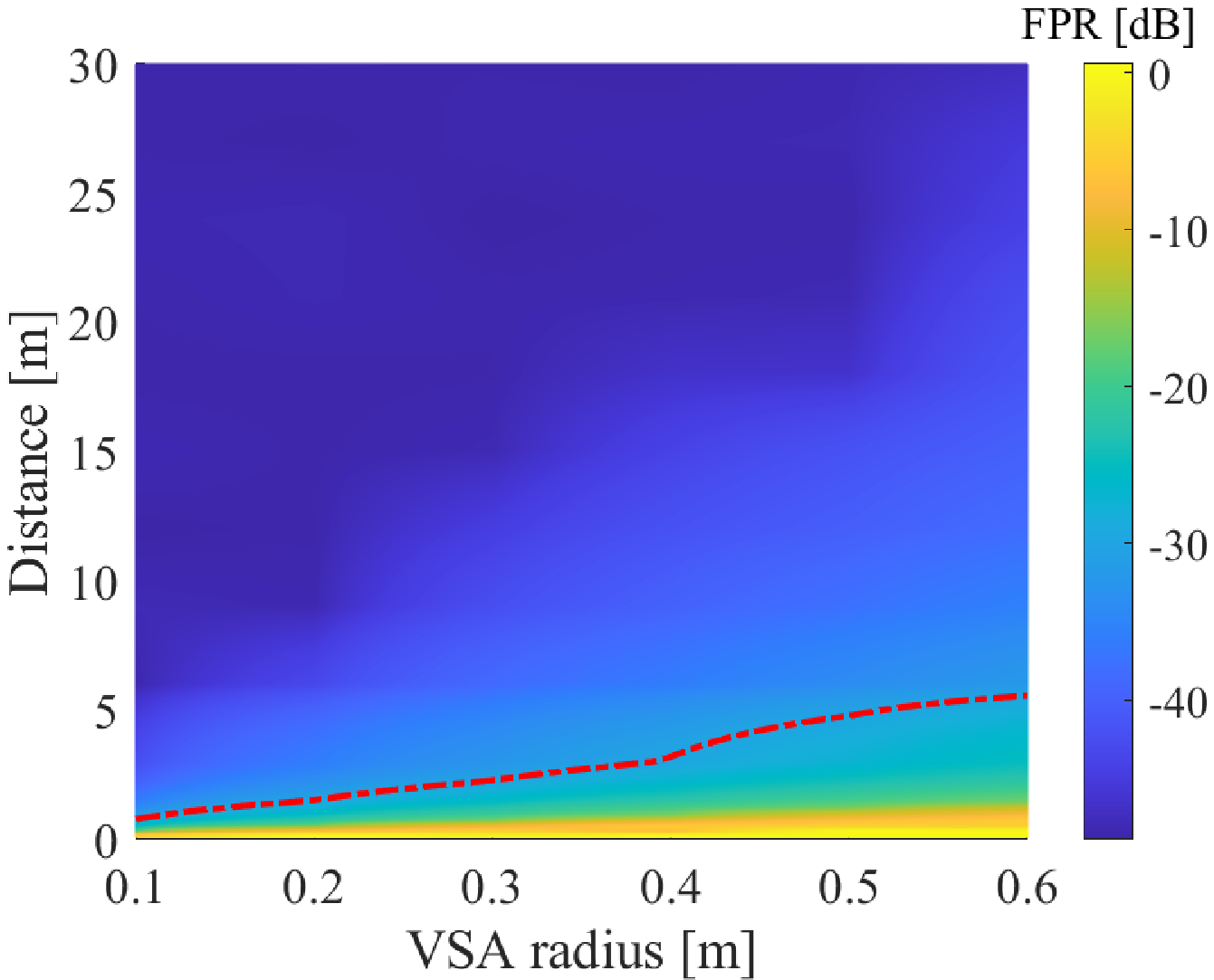}
}
\subfloat[$30^\circ$.]{
\includegraphics[width=0.24\textwidth]{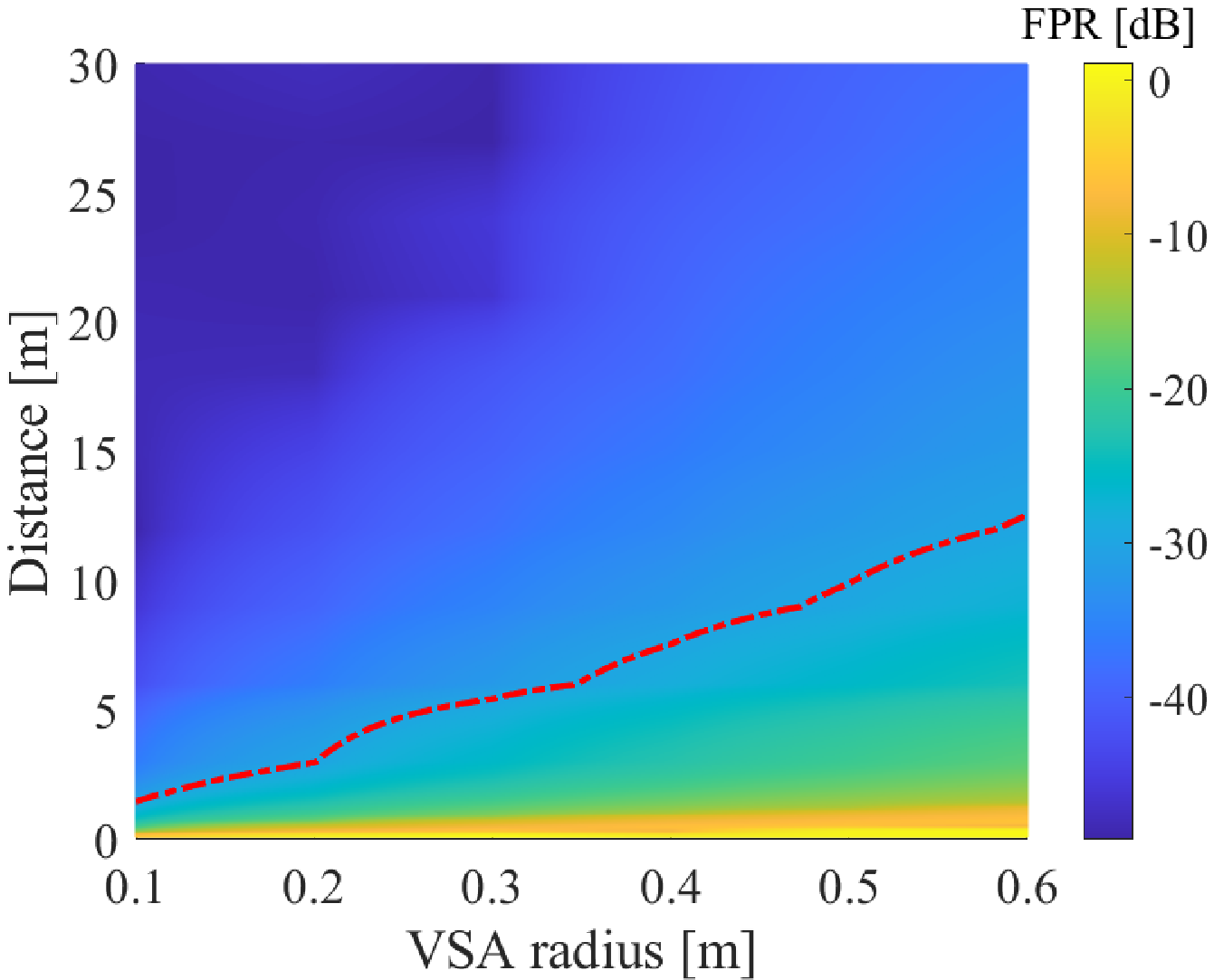}
}
\caption{{The FPR values of DSS-o-SAGE w. FFA with different antenna HPBW. The red dotted line stands for a \SI{-30}{dB} threshold.}}
\label{fig:apdvary}
\vspace{-0.5 cm}
\end{figure}
\par {Note that the above results are not sufficient to conclude that the near-field effects are not important in the DSS-o-SAGE algorithm. Since the near-field effects are related to the array aperture, distance, and the HPBW of the antenna elements, extensive simulations are conducted with different values of these parameters to observe the effects of near-fields on the estimation performance of DSS-o-SAGE algorithm. To investigate the influences of HPBW of antennas, the normalized antenna gain pattern with different HPBW is approximated by using the Gaussian main beam model~\cite{Nie2021Channel}, expressed as
\begin{equation}
    G(\varphi,\theta)=\exp{\left(-\frac{4\ln{(2)}\varphi^2}{\varphi^2_\text{3dB}}\right)}\exp{\left(\frac{4\ln{(2)}\theta^2}{\theta^2_\text{3dB}}\right)}
\end{equation}
where $\varphi_\text{3dB}$ and $\theta_\text{3dB}$ are the HPBW in azimuth and elevation planes, respectively, which are consistent in this work.}
\par {The FPR values of DSS-o-SAGE w. FFA are shown in Fig.~\ref{fig:apdvary}. To clearly see the FPR values caused by FFA, the SNR is set as \SI{60}{dB} in these simulations, for which the FPR values are lower bounded by around \SI{-45}{dB} due to noise effects. First, the FPR values generally increase as the radius of the VSA increases. As the scatter distance increases, the FPR values decrease. This is intuitively right since smaller scatter distance and larger VSA radius result in severe near-field effects. Second, the FPR values do not increase monotonically as the antenna HPBW increases. Instead, the FPR values increase when the antenna HPBW increases from $5^\circ$ to $10^\circ$ and then decrease as the HPBW grows. The reason behind this is two-fold. On one hand, the effective array aperture, defined as the largest distance between antenna elements with significant received power, increases as the HPBW increases, resulting in more degradation of estimation performance and thus higher FPR values of the DSS-o-SAGE w. FFA. On the other hand, by approximating the SWF with PWF, the receiving directions of the directional antennas in different scanning directions are incorrectly calculated. When the antenna HPBW is smaller, such a deviation of antenna receiving angles results in a larger miscalculation of the antenna gains. As a result, the signal components can not be completely subtracted, leaving strong fake paths. Third, by using a \SI{-30}{dB} FPR threshold, the distance boundary is illustrated in Fig.~\ref{fig:apdvary} with red dotted lines. The boundary generally follows a linear relationship with the VSA radius and an almost logarithmic relationship with the antenna HPBW, which is modeled by
\begin{equation}
    d_\text{th}=a\left|\ln\left(\frac{\theta_\text{3dB}}{b}\right)\right|^c\frac{r_\text{VSA}}{\lambda}
    \label{eq:dth}
\end{equation}
where $\theta_\text{3dB}$ and $r_\text{VSA}$ stands for the antenna HPBW and VSA radius. Moreover, the fitting parameters are given as $a=0.0184, b=10.773^\circ, c=1.4597$, respectively.}
\section{Estimation and Channel Characterization Results Using Real Measured Data}
\label{sec:char}
\par In this section, the estimation performance of the proposed DSS-o-SAGE algorithm and the widely used noise-elimination method is evaluated and compared using real measured data in an indoor corridor scenario, as reported in~\cite{li2022channel}. Based on the estimated channel parameters, propagation analysis is performed to reveal the propagation mechanism of THz waves. Furthermore, channel characteristics are calculated and discussed, including path loss, delay and angular spreads. 
\subsection{Measurement Campaign and Parameter Setups}
\begin{figure}
    \centering
    \subfloat[Pictures of the corridor.] {
     \label{fig:corridor}     
    \includegraphics[width=0.75\columnwidth]{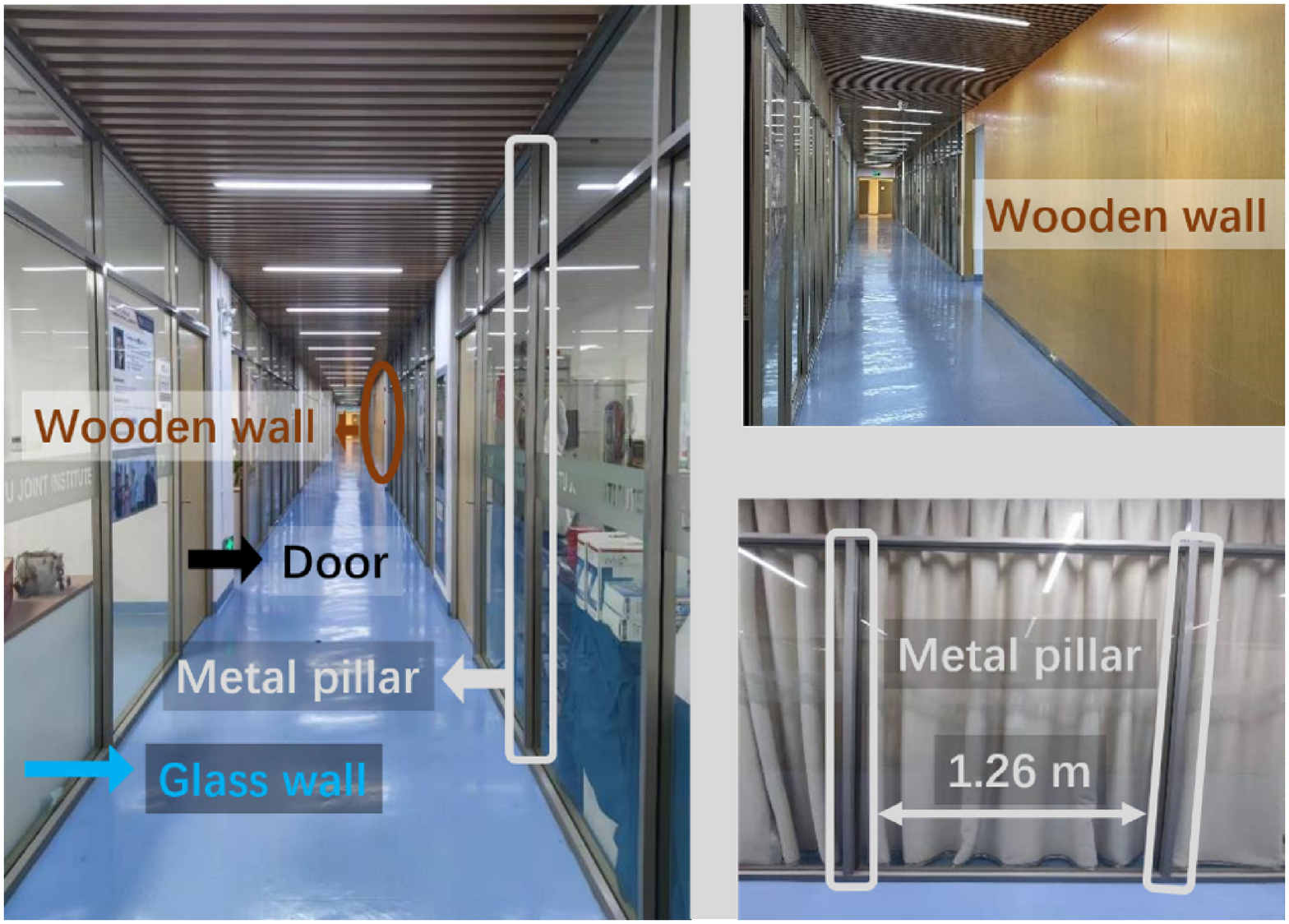}  
    }
    \quad
    \subfloat[Bird's eye view.] {
     \label{fig:deployment}     
    \includegraphics[width=0.75\columnwidth]{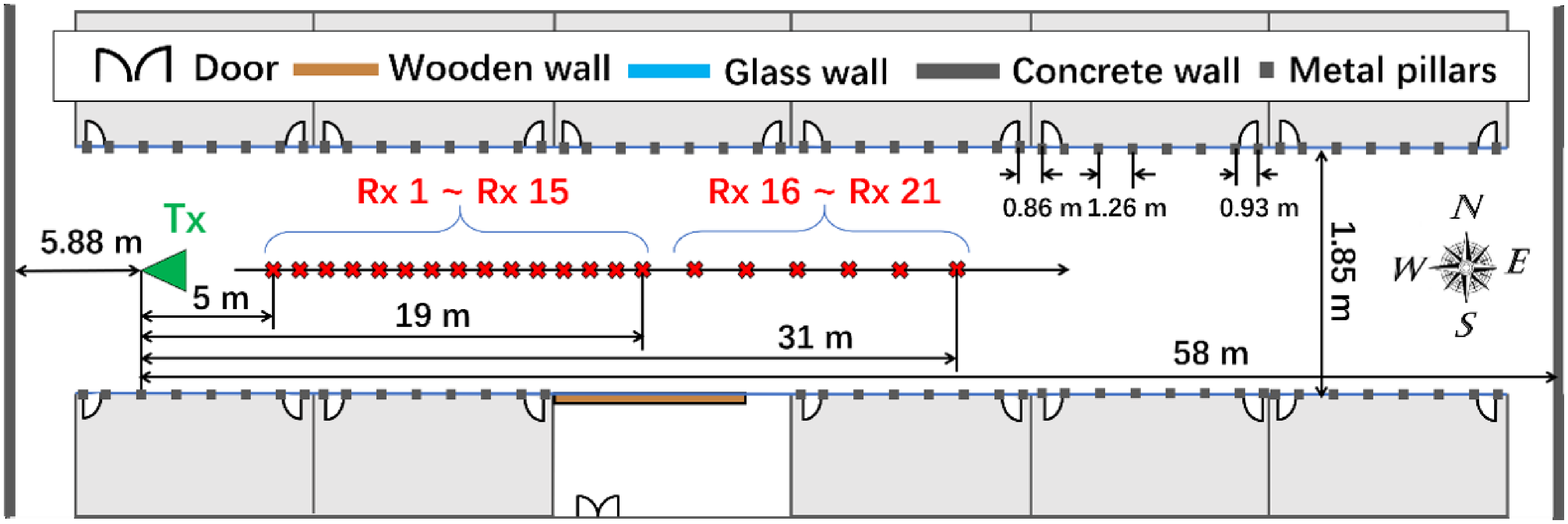}  
    }   
    \caption{The layout of the measurement campaign conducted in the corridor~\cite{li2022channel}}
    \label{fig:layout}
    \vspace{-0.5cm}
\end{figure}
\par The channel measurement campaign is conducted in \SIrange{306}{321}{GHz} in a typical indoor corridor scenario, as shown in Fig.~\ref{fig:layout}. Both sides of the corridor are furnished with glass walls, which are connected together by metal pillars. The Tx/Rx separation distance ranges from \SI{5}{m} to \SI{31}{m} with 21 receiver positions. Moreover, the Tx antenna remains static, while the Rx antenna scans the spatial domain, as $-20^\circ\sim20^\circ$ in the elevation plane and $0\sim360^\circ$ in the azimuth plane, both with a rotation step of $10^\circ$. Therefore, we estimate the DoAs of MPCs, while the DoDs of MPCs are omitted. The rotator used is the same as that mentioned in Sec.~\ref{sec:pattern}. For detailed descriptions of the channel sounders, measurement setups, and measurement deployments, please refer to~\cite{li2022channel}.
\par Based on the measured data, only MPCs with path gains larger than a certain threshold $\alpha_{\text{th}}$ are estimated, as 
\begin{equation}
\alpha_{\text{th}}=\max(\alpha^\text{FS}_{\text{LoS}}/1000),
\end{equation}
where $\alpha^\text{FS}_\text{LoS}$ denotes the free space path gain, calculated using the Friss' law, i.e., $\alpha^\text{FS}_\text{LoS}=c/4\pi fd$, with $c$, $f$, $d$ standing for the speed of light, carrier frequency and distance. Moreover, the number $1000$ indicates a \SI{30}{dB} dynamic range. 
\par For noise-elimination method, all samples of CIRs that are stronger than $\alpha_{\text{th}}$ are regarded as MPCs. For the DSS-o-SAGE algorithm, the initialization cycle starts with all MPC parameters set as zero and terminates when the estimated path gain is smaller than the threshold. The convergence of the DSS-o-SAGE algorithm is judged by the criterion in \eqref{eq:converge}, with the ratio threshold parameter set to $0.001$. {The results of DSS-o-SAGE w. FFA are obtained in a similar way to those of DSS-o-SAGE, only with the all scatter distances approaching infinity. Note that the PWF-SAGE and SWF-SAGE algorithms are omitted here since they produce significant fake MPCs and thus greatly deviated channel characteristics under the severe phase instability, as discussed in Sec.~\ref{sec:accu}.}
\begin{figure*}[!tbp]
\centering
\subfloat[Results with noise-elimination.]{
\includegraphics[width=0.24\textwidth]{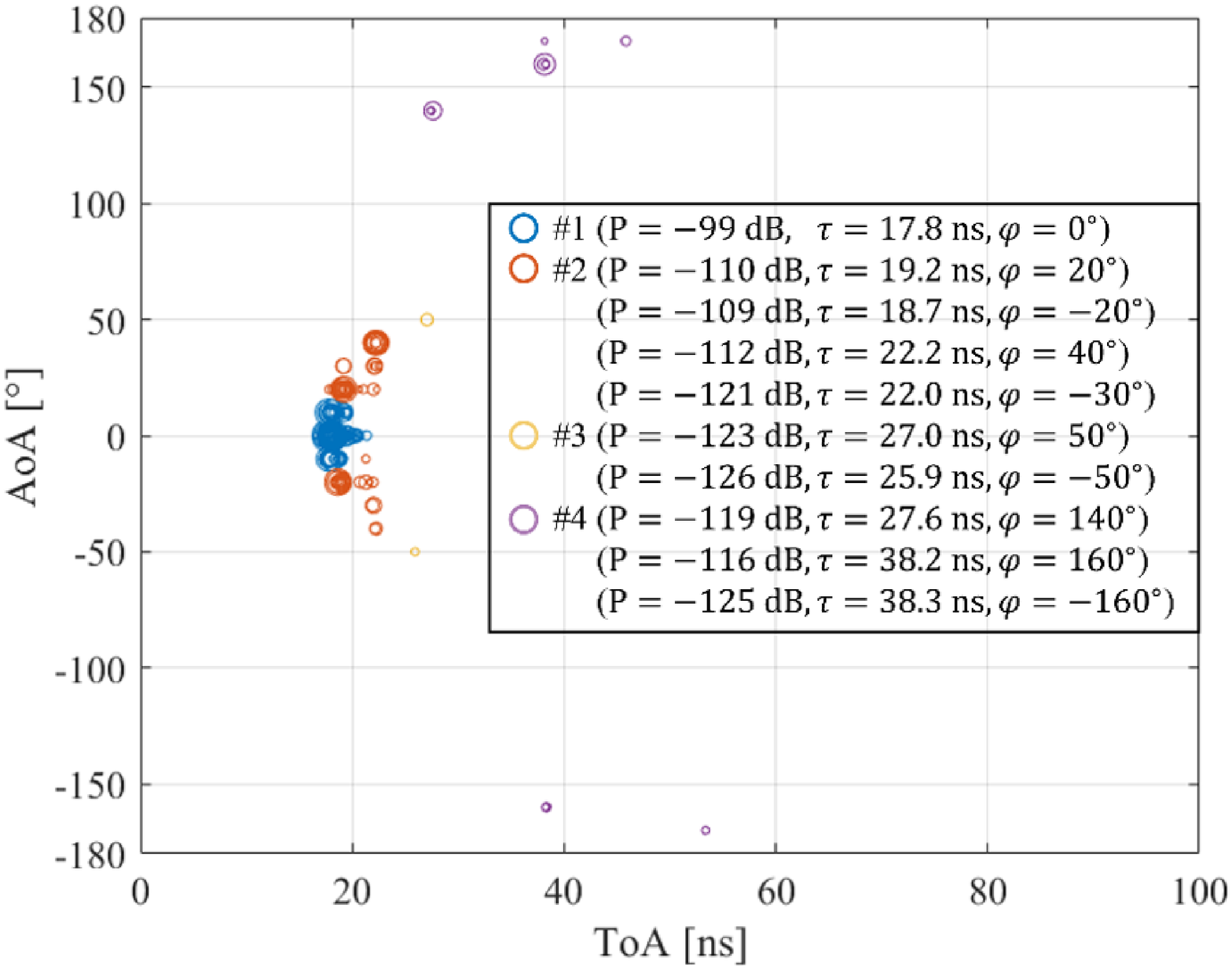}
}
\subfloat[Results with DSS-o-SAGE.]{
\includegraphics[width=0.24\textwidth]{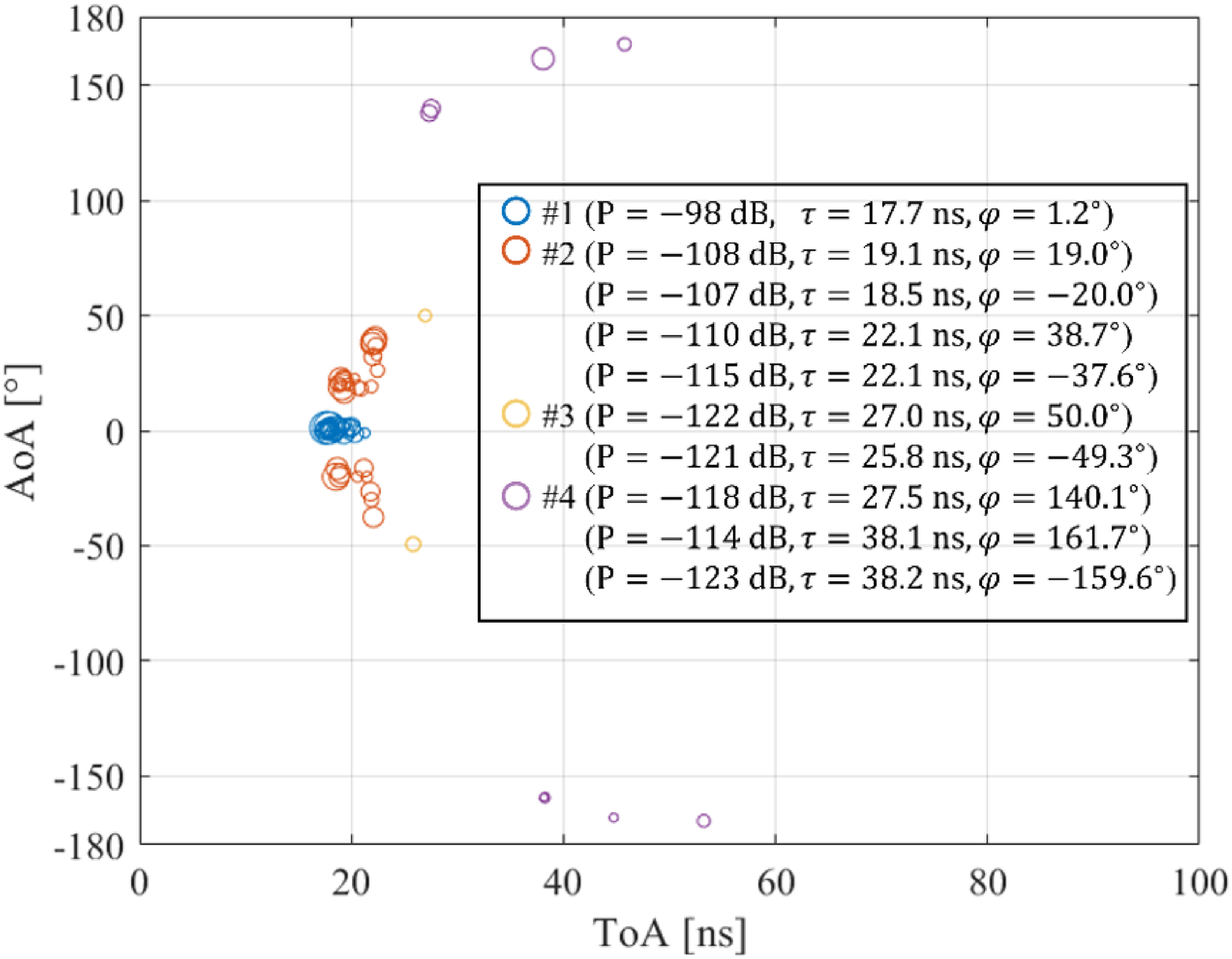}
}
\subfloat[{Results with DSS-o-SAGE w. FFA.}]{
\includegraphics[width=0.24\textwidth]{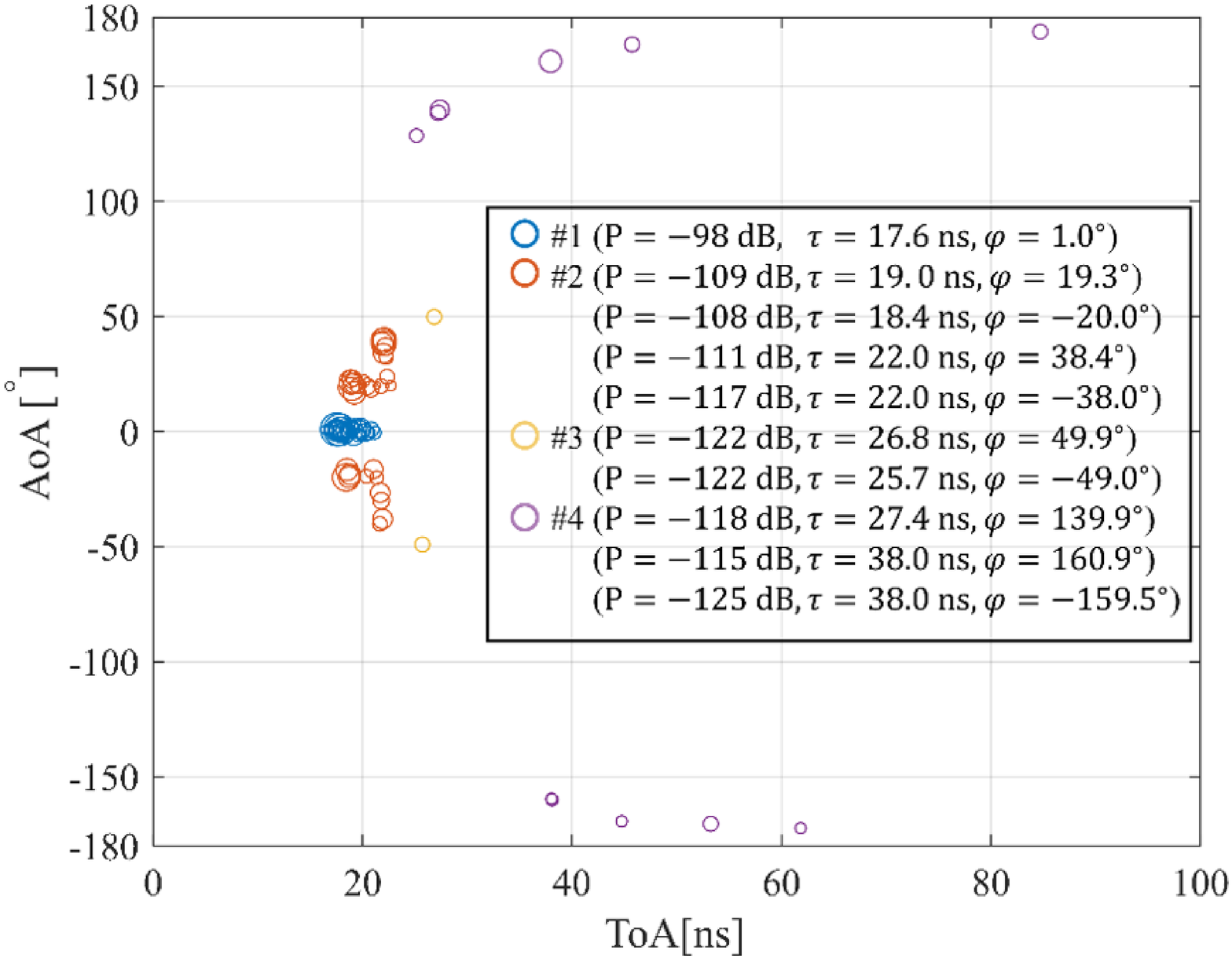}
}
\subfloat[Propagation of main MPCs.]{
\raisebox{0.2\height}{\includegraphics[width=0.2\textwidth]{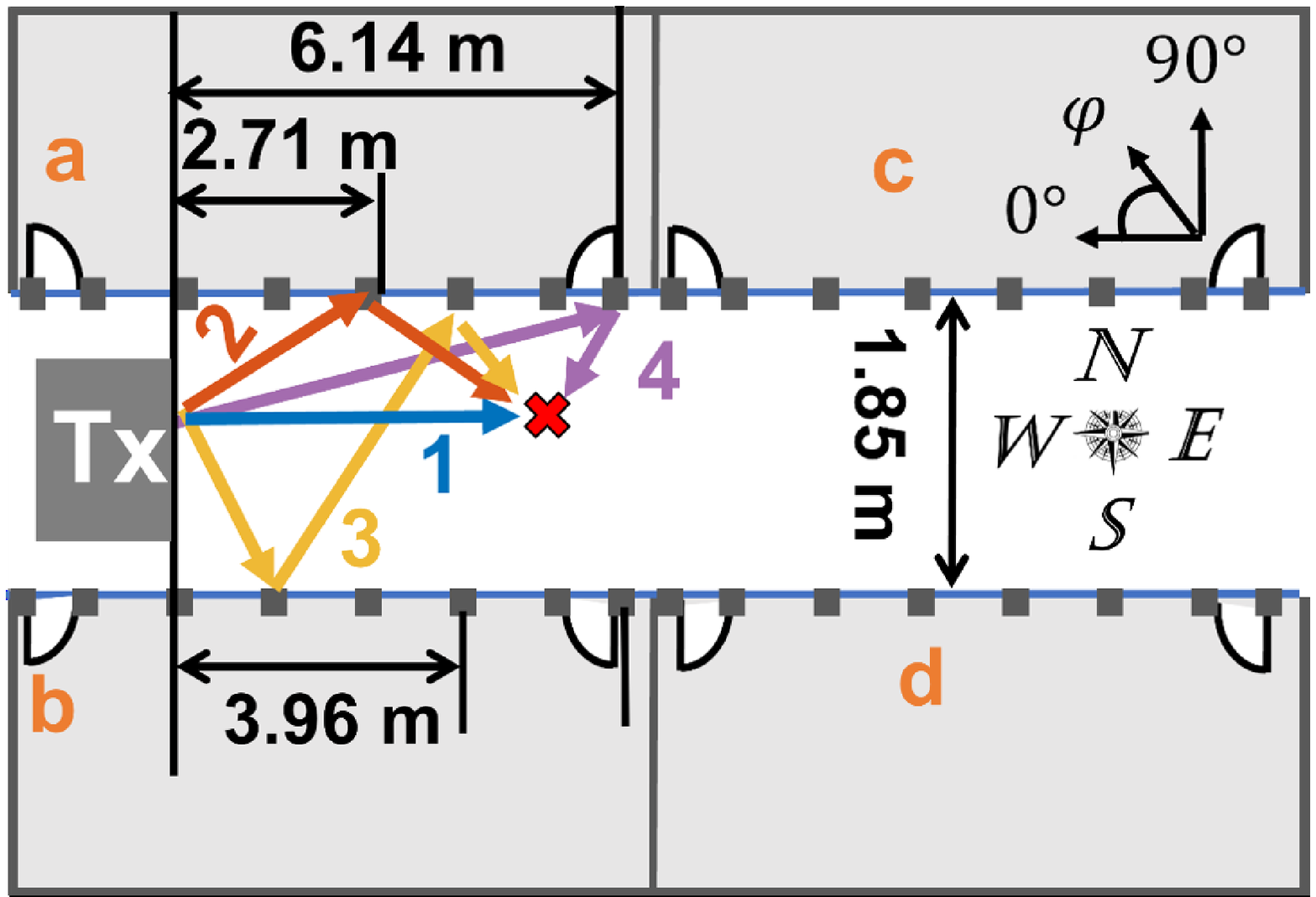}}
}
\caption{Multi-path propagation analysis based on measurement results.}
\label{fig:clustering}
\vspace{-0.3 cm}
\end{figure*}
\subsection{Propagation Analysis}
\par Based on the estimated channel parameters, the physical reason for different MPCs can be analyzed. Taking the first Rx point as an example, the estimated MPCs and corresponding propagation paths are shown in Fig.~\ref{fig:clustering}, from which several observations can be made as follows. First, judging by the parameters of the main MPCs, there are mainly four types of MPCs observed in the first Rx point, including the LoS path, the once and twice reflection paths from metal pillars in front of the Rx, as well as the back-scattering from metal pillars, as shown in Fig.~\ref{fig:clustering} (d). The LoS path dominates the measured channel, while a \SIrange{10}{20}{dB} reflection/scattering loss is observed for the NLoS path. Second, comparing the extracted MPCs by using the DSS-o-SAGE algorithm and noise-elimination method, it can be seen that the different kinds of MPCs are more separated based on the DSS-o-SAGE algorithm, since the effects of the antenna radiation pattern are eliminated. In contrast, the results with noise-elimination contains many fake MPCs, such as the duplicates of the LoS path observed in AoA at $10^\circ$ and $-10^\circ$, for which the LoS path and the reflection paths in AoA at $20^\circ$ and $-20^\circ$ are nearly connected together and hard to distinguish. Specifically, the number of MPCs by using DSS-o-SAGE algorithm and noise-elimination is 70 and 161, respectively. Third, by comparing the parameters of main MPCs listed in Fig.~\ref{fig:clustering}, it can be seen that the results with noise-elimination slightly deviates from those with DSS-o-SAGE, due to the limited temporal and spatial resolution of the noise-elimination method. Therefore, it is necessary to utilize the DSS-o-SAGE algorithm instead of the noise-elimination method to obtain more reasonable and correct results. {Fourth, the results by using DSS-o-SAGE w. FFA are very similar to those obtained using DSS-o-SAGE algorithm, with only sight difference on the estimated parameters of the MPCs. In fact, using the empirical formula in~\eqref{eq:dth}, the scatter distance threshold is calculated as \SI{0.94}{m}, which is below most of the real scatter distances and thus the near-field effects have insignificant effects on the estimation performance of DSS-o-SAGE in this case.}
\subsection{Channel Characteristics}
\par The estimated MPC parameters directly affect the calculations of channel characteristics. In this part, we calculate and analyze the channel characteristics based on different channel parameter estimation algorithms, including path loss, delay and angular spreads.
\subsubsection{Path loss}
\begin{table*}[tbp]
    \centering
    \caption{Channel characteristics calculated based on different channel parameter estimation algorithms in the corridor.}
    \includegraphics[width = 2\columnwidth]{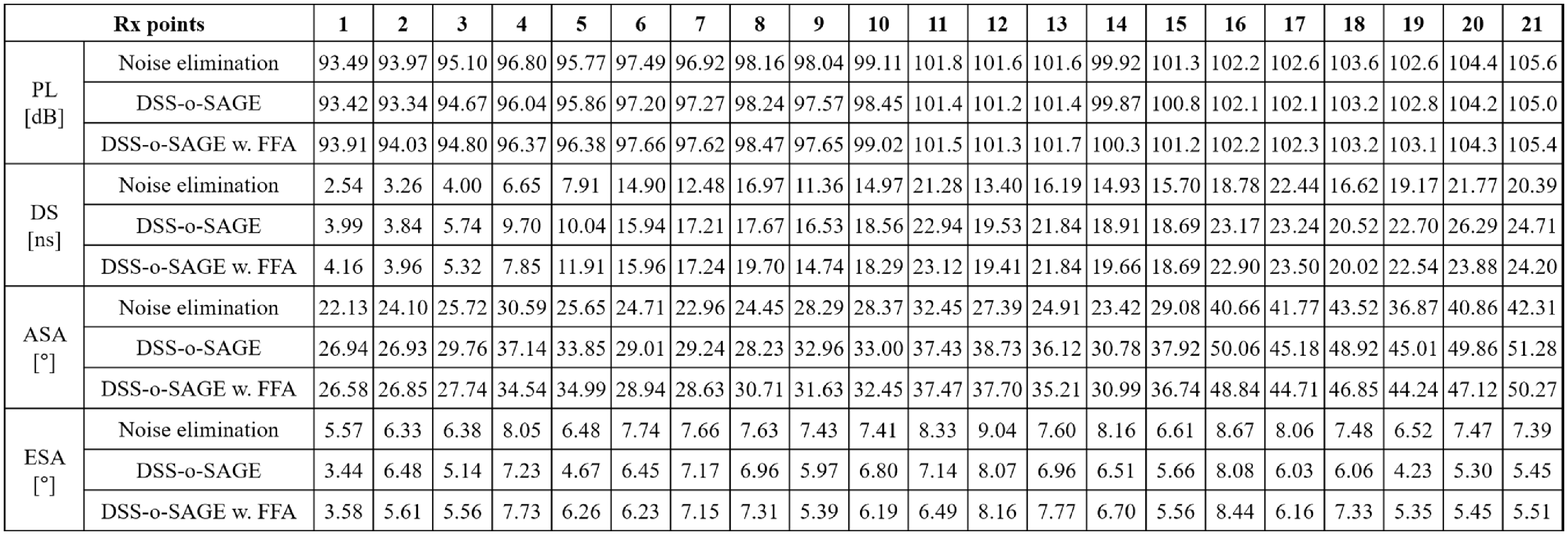}
    \label{tab:char}
    \vspace{-0.3 cm}
\end{table*}
\begin{table}[tbp]
    \centering
    \caption{Channel characteristic distribution parameters calculated based on different channel parameter estimation algorithms in the corridor.}    
    \includegraphics[width = 0.8\columnwidth]{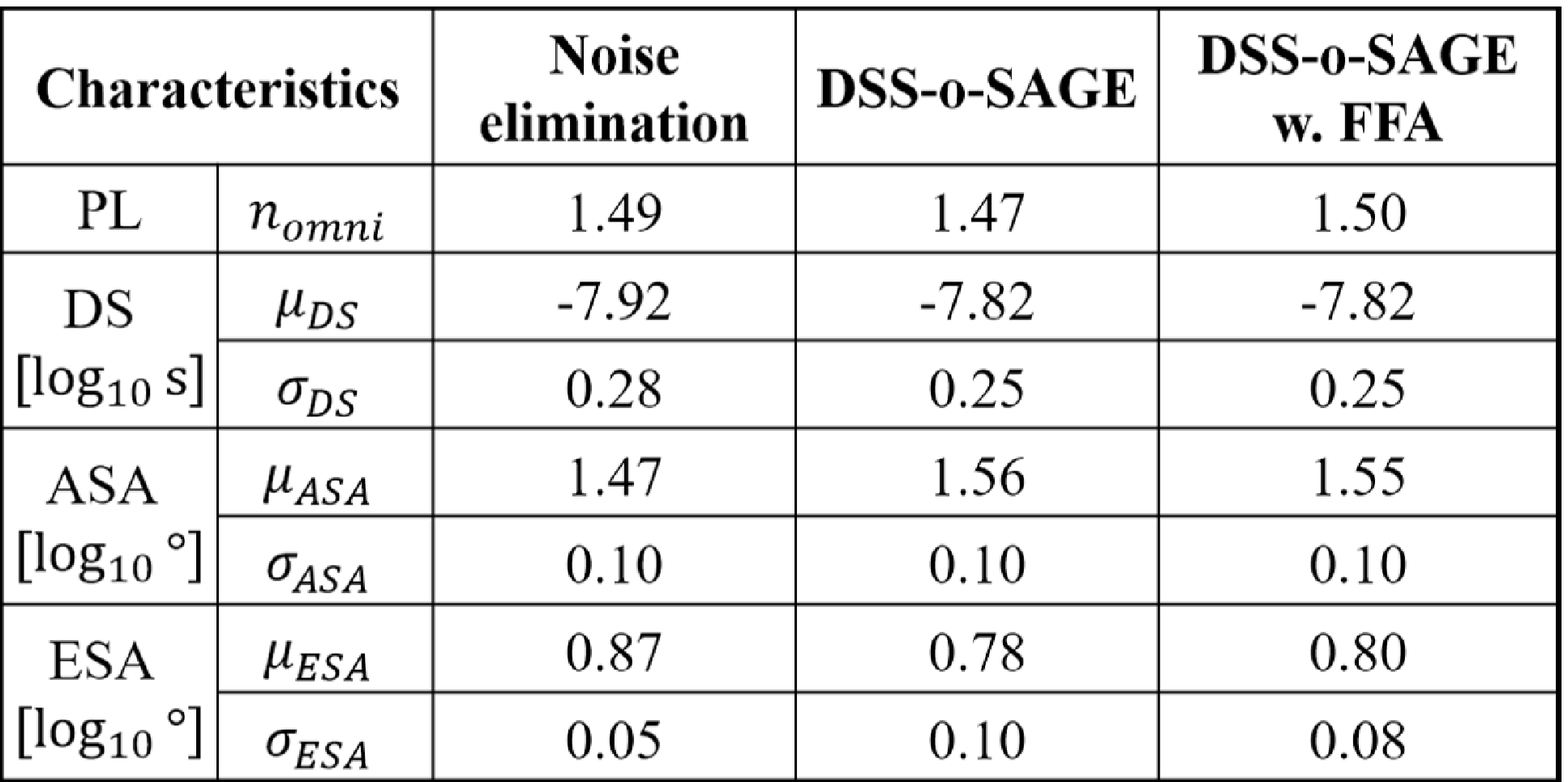}  \label{tab:chardis}
    \vspace{-0.5 cm}
\end{table}
\par The path loss, also known as large-scale fading, is of great importance for the evaluation of the link budget in the design of communication systems. Specifically, in our channel measurement campaigns, the path loss is calculated as the summation of the received power of all MPCs, as
\begin{equation}
    \text{PL}~[\text{dB}]=-10\log_{10}(\sum_{l=1}^L{\left|\alpha_{l}\right|^2}).
\end{equation}
\par Generally, the path loss is modeled as a linear function with respect to the logarithm of the Euclidean distance between Tx and Rx. In line with this, a close-in (CI) free space reference distance model is used, which is defined as
\begin{equation}
    \text{PL}^\text{CI} [\text{dB}]=10n\log_{10}\frac{d}{d_0}+\text{FSPL}(d_0)
\end{equation}
where $n$ stands for the path loss exponent (PLE) of the CI model. Additionally, $d$ is the Euclidean distance between Tx and Rx and $d_0$ stands for the reference distance, which is selected as \SI{1}{m} here. $\text{FSPL}(d_0)$ denotes the free space path loss at distance $d_0$ and is calculated as $20\log_{10}(\alpha^{\text{FS}}_{\text{LoS}}(f,d_0))$.
\par The path loss results are shown in Table~\ref{tab:char}and~\ref{tab:chardis}. First, the path loss results of noise-elimination method and DSS-o-SAGE algorithm are close. The reason is that for a certain MPC, its duplicates in all scanning directions are treated as MPCs by the noise-elimination method, and the sum of their power is close to the real power of the MPC. As a result, similar path loss results are observed for noise-elimination and DSS-o-SAGE algorithm.  {Second, the path loss results obtained by using DSS-o-SAGE w. FFA are slightly overestimated compared to the results without FFA. The reason is that by using the FFA, the path gains of MPCs are slightly underestimated due to the model mismatch between PWF and SWF. Nonetheless, the deviation is insignificant, which means that FFA can be applied to reduce the computational complexity in our measurements.} Third, with the DSS-o-SAGE algorithm, the path loss in the indoor corridor scenario at \SI{300}{GHz} is characterized with a PLE around 1.47. The small PLE attributes to the significant scattering from metal pillars, as mentioned earlier in the propagation analysis.
\subsubsection{Delay and angular spreads}
\par The power of the MPCs disperses in both temporal and spatial domains, which is evaluated by the delay and angular spreads. The delay spread, azimuth spread of arrival (ASA) and the elevation spread of arrival (ESA) are summarized in Table~\ref{tab:char}, from which we make several observations. First, for the delay spread values, the results using noise-elimination method are smaller compared to the results using DSS-o-SAGE algorithm. Second, the ASA results based on the noise-elimination method are smaller than those based on the DSS-o-SAGE algorithm, while the ESA values are larger. The reason for these observations is that by using the noise-elimination method, for a given path, significant fake MPCs with delay and DoA around the real path are obtained. As a result, the delay spread is underestimated, while the angular spreads are misleadingly close $10\sim20^\circ$ with the HPBW of the Rx antenna as $8^\circ$. Those angular spread values that are actually larger than this range tend to be smaller, while others that are expected to be smaller grow. {Third, the results show little difference whether the FFA is used or not. As shown in Table~\ref{tab:chardis}, by using the FFA in the DSS-o-SAGE algorithm, the average ASA is slightly underestimated and the average ESA is slightly overestimated, due to several fake MPCs caused by FFA. These results indicate that it is acceptable to use FFA under our measurement setups to reduce the computational complexity for channel characterization.} Fourth, by using the DSS-o-SAGE algorithm, the delay spread is averagely calculated as \SI{15.26}{ns}, while the average ASA and ESA values are $36.23^\circ$ and $6.06^\circ$, respectively. The large delay and angular spread values reveals the strong multi-path effects due to diffuse scattering from the metal pillars on both sides of the corridor.
\section{Conclusion}
\label{sec:conclude}
\par In this paper, we proposed a direction-scan sounding-oriented SAGE (DSS-o-SAGE) algorithm for channel parameter estimation in mmWave and THz bands. We introduced and validated a scanning-direction-dependent signal phase, to adequately capture the phase instability observed in mmWave and THz DSS. Based on the signal model, the DSS-o-SAGE algorithm was derived, whose computational complexity was reduced with coarse-to-fine estimation and partial data by exploiting the narrow antenna beam property of DSS. Furthermore, we evaluated the performance of the proposed algorithm and compared it with existing SAGE algorithms, considering a synthetic channel in a single path case. Influences of the phase instability and near-field, as well as the computational complexity of different algorithms, are fully investigated. Last but not least, we applied the proposed DSS-o-SAGE algorithm in real measured data, which were further compared with the widely used noise-elimination method.
\par Particularly, key observations are highlighted as follows. First, by carefully considering the phase instability of mmWave and THz DSS, the proposed DSS-o-SAGE algorithm greatly outperforms existing algorithms in terms of estimation accuracy. Second, by exploiting the narrow antenna beam property in mmWave and THz DSS, the coarse-to-fine estimation process with partial data utilized in DSS-o-SAGE algorithm greatly reduces the computational complexity. Third, due to the narrow antenna beams used in mmWave and THz DSS, the near-field effects are much weaker than one may expect from the Rayleigh distance. Specifically, the scatter distance boundary, beyond which far-field approximations can be used in DSS-o-SAGE, is linearly related to the array aperture and shows a logarithmic dependence on the antenna beam width. {In fact, under the measurement setups and scenarios that we consider, the near-field effect is not significant for channel parameter estimations and far-field approximations can be thus utilized for lower computational complexity.} Fourth, without using the DSS-o-SAGE algorithm, multiple fake MPCs from the duplicates caused by the radiation patterns of horn antennas are obtained using the noise-elimination method, which further leads to inaccurate channel characteristic results, such as smaller delay spread, smaller azimuth angle spread, and larger elevation angle spread. Based on the DSS-o-SAGE algorithm, the channels in the indoor corridor scenario at \SI{300}{GHz} are characterized, where we observe a PLE of 1.47, average delay spread of \SI{15.26}{ns}, average ASA and ESA of $36.23^\circ$ and $6.06^\circ$, respectively. In general, the proposed DSS-o-SAGE algorithm can be used to accurately estimate the MPC parameters in mmWave and THz direction-scan measurements, which is helpful for channel modeling and system design for mmWave and THz communications.
\begin{appendices}
\section{General Form of Channel Impulse Response}
\label{appen:cir}
\setcounter{equation}{0}
\renewcommand\theequation{A\arabic{equation}}
\par There are mainly three kinds of methods to measure the mmWave and THz channels, namely pulse-based time domain method, correlation-based time domain method, and VNA-based frequency domain method~\cite{han2022terahertz}. Since the pulse-based time domain method has limited measurement distance, it is usually used to measure material properties or micro-scale channels, rather than used in indoor/outdoor scenarios in DSS. In the following part, we show that for the other two methods, the measured CIR can be expressed in the same form.
\subsection{Correlation-based Method}
\par In correlation-based method, the auto-correlation function of the transmitted signal is close to a Dirac function. Therefore, after the received signal passing a correlator, the CIR is obtained, as
\begin{equation}
    \begin{split}
        h(\tau)&=\int_0^{T_c}y(t)u^*(t-t_0-\tau)dt\\
        &=\sum_{l=1}^Ls'_lR_u(\tau-\tau_{l})+w(t)
    \end{split}
\end{equation}
where $s'_l=\alpha_lc^\text{Tx}(\bm{\Omega}^\text{Tx}_{l})c^\text{Rx}(\bm{\Omega}^\text{Rx}_{l})\text{e}^{j\phi_{l}}$ denote part of the signal that is not dependent on time. Additionally $T_c$ denotes the signal duration of $u(t)$. Moreover, $R_u(\tau)=u(t)u^*(t-\tau)$ is the auto-correlation function of the transmitted signal. Moreover, $(\cdot)^*$ is the conjugate operation of complex numbers.
\par With temporal sampling, the discrete CIR is expressed as
\begin{equation}
    h[i]=\sum_{l=1}^Ls'_lR_u((i-1)\Delta\tau-\tau_{l})+w[i]
    \label{eq:corcir}
\end{equation}
\subsection{VNA-based Method}
Different from the above-mentioned two time-domain methods, the VNA-based method measures the CTFs of wireless channels and the CIRs are obtained through inverse discrete Fourier transform (IDFT). Specifically, the sampled CTF of wireless channels can be expressed as
\begin{equation}
    H[k]=\sum_{l=1}^Ls'_l\text{e}^{-j2\pi f_k\tau_{l}}+W[k]
\end{equation}
where $f_k$ denotes the carrier frequency at the $k^\text{th}$ sampling point, i.e., $f_k=f_1+(k-1)\Delta f$. $f_1$ and $f_2$ represent the lower and upper frequencies of the measured band. $\Delta f$ is the sampling interval in the frequency domain. 
\par Taking IDFT on the measured CTF, we obtain
\begin{equation}
\begin{split}
    h[i]=\sum_{l=1}^Ls'_l\text{IDFT}[\text{e}^{-j2\pi f_k\tau_{l}}]+w[i]
    \label{eq:vnadcir}
\end{split}
\end{equation}
\par Furthermore, by using the definitions of IDFT and denoting the term $\text{IDFT}[\text{e}^{-j2\pi f_k\tau_l}]$ as $v_l$, we have
\begin{subequations}
\begin{align}
    v_l[i]&=\frac{1}{K}\sum_{k=1}^K\text{e}^{-j2\pi f_k\tau_l}\text{e}^{j2\pi(k-1)(i-1)/K}\\
    &=\frac{\text{e}^{j\varphi_l}}{K}\sum_{k=1}^K\text{e}^{-j2\pi((k-1)\Delta f\tau_l-(k-1)(i-1)/K)}\\
    \label{eq:der_ukb}
    &=\frac{\text{e}^{j{\varphi}_l}}{K}\sum_{k=1}^K\text{e}^{-j2\pi((k-1)\Delta f{\tau_l}-(k-1)(i-1)\frac{\Delta\tau\Delta f}{1+\Delta\tau\Delta f})}\\
    &=\frac{\text{e}^{j{\varphi}_l}}{K}\sum_{k=1}^K\text{e}^{-j2\pi(k-1)\Delta f\Delta_{\tau_l}[i]}
\end{align}
\end{subequations}
where \eqref{eq:der_ukb} comes from the fact that $1/(K-1)=\Delta f\Delta\tau$. Besides, $\Delta_{\tau_l}[i]=\tau_l-(i-1)\Delta\tau/a$ with $a=1+\Delta\tau\Delta f$. Moreover, the additional phase term ${\varphi}_l$ is $-2\pi f_1{\tau}_l$.
\par When $\Delta f\Delta_{\tau_l}[i]=0$, it is obvious that $v_l[i]=\text{e}^{j{\varphi}_l}$. When $\Delta f\Delta_{\tau_l}[i]\neq0$, By utilizing the properties of geometric series, we obtain
\begin{subequations}
\begin{align}
        v_l[i]&=\frac{\text{e}^{j{\varphi}_l}}{K}\frac{\text{e}^{-j\pi K\Delta f\Delta_{\tau_l}[i]}}{\text{e}^{-j\pi\Delta f\Delta_{\tau_l}[i]}}\frac{\sin{(\pi K\Delta f\Delta_{\tau_l}[i])}}{\sin{(\pi\Delta f\Delta_{\tau_l}[i])}}\\
        &\triangleq\text{e}^{j\varphi'_{l}}g_{K}(\pi\Delta f\Delta_{\tau_l}[i])
        \label{eq:der_ukd}
\end{align}
\end{subequations}
where the phase term is ${\varphi}'_{l}={\varphi}_l-\pi (K-1)\Delta f\Delta_{\tau_l}[i]$. Moreover, $g_{K}(x)$ is of the form as
\begin{equation}
    g_{K}(x)=
    \begin{cases}
    1,&x=0\\
    \frac{\sin(Kx)}{K\sin(x)},&x\neq0
    \end{cases}
\end{equation}
\par Utilizing \eqref{eq:der_ukd} in \eqref{eq:vnadcir}, we have
\begin{equation}
\begin{split}
    h[i]=\sum_{l=1}^Ls'_lg_{K}(\pi\Delta f\Delta_{\tau_l}[i])\text{e}^{j\varphi'_{l}}+w'[i]
    \label{eq:vnacir}
\end{split}
\end{equation}
\subsection{Summary}
\par The measured CIRs with different methods in~\eqref{eq:corcir} and~\eqref{eq:vnacir} can be expressed with a general form, as
\begin{equation}
h[i]=\sum_{l=1}^Ls'_lR_{\tau_{l}}[i]+w[i]
\end{equation}
where the function $R_{\tau_{l}}[i]$ is of the form as
\begin{equation}
\begin{split}
    R&_{\tau_{l}}[i]=
    \begin{cases}
    R_u((i-1)\Delta\tau-\tau_l),&\text{correlation-based}\\
    g_{K}(\pi\Delta f\Delta_{\tau_l}[i])\text{e}^{j\varphi'_{l}},&\text{VNA-based}
    \end{cases}
    \label{eq:rfunction}
\end{split}
\end{equation}
\end{appendices}
\bibliographystyle{IEEEtran}
\bibliography{IEEEabrv,yuanbo}

\begin{thebibliography}{10}
\providecommand{\url}[1]{#1}
\csname url@samestyle\endcsname
\providecommand{\newblock}{\relax}
\providecommand{\bibinfo}[2]{#2}
\providecommand{\BIBentrySTDinterwordspacing}{\spaceskip=0pt\relax}
\providecommand{\BIBentryALTinterwordstretchfactor}{4}
\providecommand{\BIBentryALTinterwordspacing}{\spaceskip=\fontdimen2\font plus
\BIBentryALTinterwordstretchfactor\fontdimen3\font minus
  \fontdimen4\font\relax}
\providecommand{\BIBforeignlanguage}[2]{{%
\expandafter\ifx\csname l@#1\endcsname\relax
\typeout{** WARNING: IEEEtran.bst: No hyphenation pattern has been}%
\typeout{** loaded for the language `#1'. Using the pattern for}%
\typeout{** the default language instead.}%
\else
\language=\csname l@#1\endcsname
\fi
#2}}
\providecommand{\BIBdecl}{\relax}
\BIBdecl

\bibitem{chen2021terahertz}
Z.~Chen, C.~Han, Y.~Wu, L.~Li, C.~Huang, Z.~Zhang, G.~Wang, and W.~Tong,
  ``Terahertz wireless communications for 2030 and beyond: A cutting-edge
  frontier,'' \emph{{IEEE} Commun. Mag.}, vol.~59, no.~11, pp. 66--72, 2021.

\bibitem{xiao2017mmwave}
M.~Xiao, S.~Mumtaz, Y.~Huang, L.~Dai, Y.~Li, M.~Matthaiou, G.~K. Karagiannidis,
  E.~Björnson, K.~Yang, C.-L. I, and A.~Ghosh, ``Millimeter wave
  communications for future mobile networks,'' \emph{{IEEE} J. Sel. Areas
  Commun.}, vol.~35, no.~9, pp. 1909--1935, 2017.

\bibitem{akyildiz2018combating}
I.~F. Akyildiz, C.~Han, and S.~Nie, ``Combating the distance problem in the
  millimeter wave and terahertz frequency bands,'' \emph{{IEEE} Commun. Mag.},
  vol.~56, no.~6, pp. 102--108, 2018.

\bibitem{rappaport2019wireless}
T.~S. Rappaport, Y.~Xing, O.~Kanhere, S.~Ju, A.~Madanayake, S.~Mandal,
  A.~Alkhateeb, and G.~C. Trichopoulos, ``{Wireless communications and
  applications above 100 GHz: Opportunities and challenges for 6G and
  beyond},'' \emph{{IEEE} Access}, vol.~7, pp. 78\,729--78\,757, 2019.

\bibitem{Erden202028}
F.~Erden, O.~Ozdemir, and I.~Guvenc, ``{28 GHz mmWave Channel Measurements and
  Modeling in a Library Environment},'' \emph{in Proc. of IEEE RWS}, pp.
  52--55, 2020.

\bibitem{Rappaport2015Wideband}
T.~S. Rappaport, G.~R. MacCartney, M.~K. Samimi, and S.~Sun, ``Wideband
  millimeter-wave propagation measurements and channel models for future
  wireless communication system design,'' \emph{{IEEE} Trans. Commun.},
  vol.~63, no.~9, pp. 3029--3056, 2015.

\bibitem{Xing2021mmwave}
Y.~Xing, T.~S. Rappaport, and A.~Ghosh, ``{Millimeter Wave and Sub-THz Indoor
  Radio Propagation Channel Measurements, Models, and Comparisons in an Office
  Environment},'' \emph{{IEEE} Commun. Lett.}, vol.~25, no.~10, pp. 3151--3155,
  2021.

\bibitem{ju2021millimeter}
S.~Ju, Y.~Xing, O.~Kanhere, and T.~S. Rappaport, ``Millimeter wave and
  sub-terahertz spatial statistical channel model for an indoor office
  building,'' \emph{{IEEE} J. Sel. Areas Commun.}, vol.~39, no.~6, pp.
  1561--1575, 2021.

\bibitem{yi2021channel}
Y.~Chen, Y.~Li, C.~Han, Z.~Yu, and G.~Wang, ``{Channel Measurement and
  Ray-Tracing-Statistical Hybrid Modeling for Low-Terahertz Indoor
  Communications},'' \emph{{IEEE} Trans. Wireless Commun.}, vol.~20, no.~12,
  pp. 8163--8176, 2021.

\bibitem{cheng2019characterization}
C.-L. Cheng and A.~Zaji{\'c}, ``{Characterization of propagation phenomena
  relevant for 300 GHz wireless data center links},'' \emph{{IEEE} Trans.
  Antennas Propag.}, vol.~68, no.~2, pp. 1074--1087, 2019.

\bibitem{priebe2011channel}
S.~Priebe, C.~Jastrow, M.~Jacob, T.~Kleine-Ostmann, T.~Schrader, and
  T.~K{\"u}rner, ``{Channel and propagation measurements at 300 GHz},''
  \emph{{IEEE} Trans. Antennas Propag.}, vol.~59, no.~5, pp. 1688--1698, 2011.

\bibitem{he2021channel}
J.~He, Y.~Chen, Y.~Wang, Z.~Yu, and C.~Han, ``{Channel Measurement and
  Path-Loss Characterization for Low-Terahertz Indoor Scenarios},'' \emph{in
  Proc. of IEEE ICC Workshops}, pp. 1--6, 2021.

\bibitem{li2022channel}
Y.~Li, Y.~Wang, Y.~Chen, Z.~Yu, and C.~Han, ``{Channel Measurement and Analysis
  in an Indoor Corridor Scenario at 300 GHz},'' \emph{in Proc. of IEEE ICC},
  2022.

\bibitem{abbasi2023thz}
N.~A. Abbasi, J.~L. Gomez, R.~Kondaveti, S.~M. Shaikbepari, S.~Rao,
  S.~Abu-Surra, G.~Xu, J.~Zhang, and A.~F. Molisch, ``{THz Band Channel
  Measurements and Statistical Modeling for Urban D2D Environments},''
  \emph{IEEE Transactions on Wireless Communications}, vol.~22, no.~3, pp.
  1466--1479, 2023.

\bibitem{eckhardt2021channel}
J.~M. Eckhardt, V.~Petrov, D.~Moltchanov, Y.~Koucheryavy, and T.~K{\"u}rner,
  ``{Channel Measurements and Modeling for Low-Terahertz Band Vehicular
  Communications},'' \emph{{IEEE} J. Sel. Areas Commun.}, vol.~39, no.~6, pp.
  1590--1603, 2021.

\bibitem{Lyu2023Measurement}
Y.~Lyu, Z.~Yuan, H.~Gao, Q.~Zhu, X.~Zhang, and W.~Fan, ``{Measurement-based
  channel characterization in a large hall scenario at 300 GHz},'' \emph{China
  Commun.}, vol.~20, no.~4, pp. 118--131, 2023.

\bibitem{han2022terahertz}
C.~Han, Y.~Wang, Y.~Li, Y.~Chen, N.~A. Abbasi, T.~Kürner, and A.~F.~Molisch,
  ``Terahertz wireless channels: A holistic survey on measurement, modeling,
  and analysis,'' \emph{{IEEE} Commun. Surveys Tuts.}, vol.~24, no.~3, pp.
  1670--1707, 2022.

\bibitem{capon1969high}
J.~Capon, ``High-resolution frequency-wavenumber spectrum analysis,''
  \emph{Proceedings of the IEEE}, vol.~57, no.~8, pp. 1408--1418, 1969.

\bibitem{bartlett1948smoothing}
M.~S. Bartlett, ``Smoothing periodograms from time-series with continuous
  spectra,'' \emph{Nature}, vol. 161, no. 4096, pp. 686--687, 1948.

\bibitem{Zhang2017Channel}
J.~Zhang and M.~Haardt, ``{Channel estimation for hybrid multi-carrier mmwave
  MIMO systems using three-dimensional unitary esprit in DFT beamspace},''
  \emph{IEEE International Workshop on CAMSAP}, pp. 1--5, 2017.

\bibitem{Guo2017Millimeter}
Z.~Guo, X.~Wang, and W.~Heng, ``Millimeter-wave channel estimation based on 2-d
  beamspace music method,'' \emph{{IEEE} Trans. Wireless Commun.}, vol.~16,
  no.~8, pp. 5384--5394, 2017.

\bibitem{thoma2004rimax}
R.~Thom{\"a}, M.~Landmann, and A.~Richter, ``{RIMAX—A maximum likelihood
  framework for parameter estimation in multidimensional channel sounding},''
  \emph{in Proc. of ISAP}, pp. 53--56, 2004.

\bibitem{moon1996expectation}
T.~K. Moon, ``The expectation-maximization algorithm,'' \emph{{IEEE} Signal
  Process. Mag.}, vol.~13, no.~6, pp. 47--60, 1996.

\bibitem{Fessler1994Space}
J.~Fessler and A.~Hero, ``Space-alternating generalized
  expectation-maximization algorithm,'' \emph{{IEEE} Trans. Signal Process.},
  vol.~42, no.~10, pp. 2664--2677, 1994.

\bibitem{fleury1999channel}
B.~H. Fleury, M.~Tschudin, R.~Heddergott, D.~Dahlhaus, and K.~I. Pedersen,
  ``{Channel parameter estimation in mobile radio environments using the SAGE
  algorithm},'' \emph{{IEEE} J. Sel. Areas Commun.}, vol.~17, no.~3, pp.
  434--450, 1999.

\bibitem{Shutin2011sparse}
D.~Shutin and B.~H. Fleury, ``{Sparse Variational Bayesian SAGE Algorithm With
  Application to the Estimation of Multipath Wireless Channels},'' \emph{{IEEE}
  Trans. Signal Process.}, vol.~59, no.~8, pp. 3609--3623, 2011.

\bibitem{ou2016sage}
L.~Ouyang and X.~Yin, ``{A SAGE algorithm for channel estimation using signal
  eigenvectors for direction-scan sounding},'' \emph{in Proc. of IEEE PIMRC},
  pp. 1--6, 2016.

\bibitem{yin2016performance}
X.~Yin, L.~Ouyang, and H.~Wang, ``{Performance comparison of SAGE and MUSIC for
  channel estimation in direction-scan measurements},'' \emph{{IEEE} Access},
  vol.~4, pp. 1163--1174, 2016.

\bibitem{yin2017scatter}
X.~Yin, S.~Wang, N.~Zhang, and B.~Ai, ``Scatterer localization using
  large-scale antenna arrays based on a spherical wave-front parametric
  model,'' \emph{{IEEE} Trans. Wireless Commun.}, vol.~16, no.~10, pp.
  6543--6556, 2017.

\bibitem{jiang2022arma}
S.~Jiang, W.~Wang, T.~Jost, P.~Peng, and Y.~Sun, ``{An ARMA-Filter Based SAGE
  Algorithm for Ranging in Diffuse Scattering Environment},'' \emph{{IEEE}
  Trans. Veh. Technol.}, vol.~71, no.~3, pp. 3361--3366, 2022.

\bibitem{hong2023joint}
J.~Hong, J.~Rodríguez-Piñeiro, X.~Yin, and Z.~Yu, ``Joint channel parameter
  estimation and scatterers localization,'' \emph{{IEEE} Trans. Wireless
  Commun.}, vol.~22, no.~5, pp. 3324--3340, 2023.

\bibitem{Zhou2023Novel}
Z.~Zhou, C.-X. Wang, L.~Zhang, J.~Huang, L.~Xin, E.-H.~M. Aggoune, and Y.~Miao,
  ``{A Novel SAGE Algorithm for Estimating Parameters of Wideband Spatial
  Nonstationary Wireless Channels With Antenna Polarization},'' \emph{IEEE
  Transactions on Antennas and Propagation}, vol.~71, no.~9, pp. 7457--7472,
  2023.

\bibitem{li2023Antenna}
M.~Li, F.~Zhang, Y.~Lyu, Z.~Yuan, and W.~Fan, ``Antenna de-embedding in
  directional channel measurements with virtual array concept and experimental
  validation,'' \emph{{IEEE} Trans. Antennas Propag.}, pp. 1--1, 2023.

\bibitem{Parssinen20216gwhitepaper}
A.~Pärssinen, M.~Alouini, M.~Berg, T.~Kuerner, P.~Kyösti, M.~E. Leinonen,
  M.~Matinmikko-Blue, E.~McCune, U.~Pfeiffer, and P.~Wambacq, ``{White paper on
  RF enabling 6G opportunities and challenges from technology to spectrum},''
  \emph{6G Flagship Ecosystem}, {Apr.} 2021, [Online]. Available:
  \url{https://www.6gchannel.com/items/6g-white-paper-rf-spectrum/}.

\bibitem{Lyu2021Design}
Y.~Lyu, A.~W. Mbugua, K.~Olesen, P.~Kyösti, and W.~Fan, ``{Design and
  Validation of the Phase-Compensated Long-Range Sub-THz VNA-Based Channel
  Sounder},'' \emph{{IEEE} Antennas Wireless Propag. Lett.}, vol.~20, no.~12,
  pp. 2461--2465, 2021.

\bibitem{Lyu2023Enabling}
Y.~Lyu, Z.~Yuan, M.~Li, A.~W. Mbugua, P.~Kyösti, and W.~Fan, ``Enabling
  long-range large-scale channel sounding at sub-thz bands: Virtual array and
  radio-over-fiber concepts,'' \emph{{IEEE} Commun. Mag.}, vol.~62, no.~2, pp.
  16--22, 2024.

\bibitem{Lyu2023Virtual}
Y.~Lyu, Z.~Yuan, F.~Zhang, P.~Kyösti, and W.~Fan, ``{Virtual Antenna Array for
  W-Band Channel Sounding: Design, Implementation, and Experimental
  Validation},'' \emph{{IEEE} J. Sel. Topics Signal Process.}, vol.~17, no.~4,
  pp. 729--744, 2023.

\bibitem{Cai2020Trajectory}
X.~Cai, W.~Fan, X.~Yin, and G.~F. Pedersen, ``Trajectory-aided
  maximum-likelihood algorithm for channel parameter estimation in
  ultrawideband large-scale arrays,'' \emph{{IEEE} Trans. Antennas Propag.},
  vol.~68, no.~10, pp. 7131--7143, 2020.

\bibitem{li2022Virtual}
M.~Li, F.~Zhang, Y.~Ji, and W.~Fan, ``Virtual antenna array with directional
  antennas for millimeter-wave channel characterization,'' \emph{{IEEE} Trans.
  Antennas Propag.}, vol.~70, no.~8, pp. 6992--7003, 2022.

\bibitem{yuan2023Millimeter}
Z.~Yuan, F.~Zhang, Y.~Zhang, J.~Zhang, G.~F. Pedersen, and W.~Fan, ``On phase
  mode selection in the frequency-invariant beamformer for near-field mmwave
  channel characterization,'' \emph{{IEEE} Trans. Antennas Propag.}, vol.~71,
  no.~11, pp. 8975--8986, 2023.

\bibitem{7744514}
Y.~Ji, W.~Fan, and G.~F. Pedersen, ``Near-field signal model for large-scale
  uniform circular array and its experimental validation,'' \emph{{IEEE}
  Antennas Wireless Propag. Lett.}, vol.~16, pp. 1237--1240, 2017.

\bibitem{Richmond2015Parameter}
C.~D. Richmond and L.~L. Horowitz, ``Parameter bounds on estimation accuracy
  under model misspecification,'' \emph{{IEEE} Trans. Signal Process.},
  vol.~63, no.~9, pp. 2263--2278, 2015.

\bibitem{Nie2021Channel}
S.~Nie and I.~F. Akyildiz, ``Channel modeling and analysis of
  inter-small-satellite links in terahertz band space networks,'' \emph{{IEEE}
  Trans. Commun.}, vol.~69, no.~12, pp. 8585--8599, 2021.

\end{thebibliography}
\end{document}